\documentclass[fleqn,10pt]{wlscirep}
\usepackage[utf8]{inputenc}
\usepackage[T1]{fontenc}
\usepackage{mathrsfs}

\usepackage{longtable}
\usepackage{geometry}
\usepackage{amsmath}
\usepackage{tikz}
\makeatletter

\usepackage{calrsfs}
\DeclareMathAlphabet{\pazocal}{OMS}{zplm}{m}{n}

\title{Non-Pharmaceutical Interventions Reshape Network Immunization Outcomes}

\author[1,2]{Sámuel G. Balogh}
\author[1,3]{Gergely Ódor}
\author[1,4,*]{Márton Karsai}

\affil[1]{National Laboratory of Health Security, HUN-REN Alfréd Rényi Institute of Mathematics, Budapest, 1053, Hungary}
\affil[2]{Faculty of Electrical Engineering, Mathematics and Computer Science, Delft University of Technology, 2600 GA Delft, The Netherlands}
\affil[3]{Institute for Hygiene and Applied Immunology, Medical University of Vienna, Vienna, 1090, Austria}
\affil[4]{Department of Network and Data Science, Central European University, Vienna, 1100, Austria}

\affil[*]{karsaim@ceu.edu}


\begin{abstract}
Herd immunity is shaped not only by the infection capacity of a spreading epidemic or the contact structure of the hosting population, but also by how and under what circumstances individuals acquire immunity. Immunization strategies may interact with ongoing non-pharmaceutical interventions, which commonly aim to reduce social contact numbers. We demonstrate that these interactions can induce unexpectedly strong and counterintuitive effects on herd immunity. We explore these phenomena on spatially embedded contact networks and uncover a reversal in the relative effectiveness of disease- versus vaccine-induced immunization schemes, highlighting the average number of contacts as a critical determinant of emerging herd immunity. In sparse geometric networks with limited degree heterogeneity, uniform vaccination proves most effective; however, as average contact numbers increase, naturally acquired immunity ultimately becomes the better strategy. We show that this phenomenon may emerge not only in synthetic networks but also in real-world mixing networks, observed during non-pharmaceutical intervention periods across multiple states of the United States.
\end{abstract}
\begin{document}

\flushbottom
\maketitle
\thispagestyle{empty}

\section*{Introduction}

During an unfolding pandemic, countries may adopt different intervention strategies to balance between epidemic control, societal consequences, and economic burden~\cite{Brisson2003EconomicHerdImmunity, Ash2022_economic_disease, Dobson2023_economic_balancing}. Throughout the COVID-19 pandemic, several non-pharmaceutical interventions have been applied~\cite{perra2021non}, ranging from strict measures, such as national lock-downs, school closings and travel bans, to weaker mitigation strategies like mask-wearing and social distancing. All these measures essentially aim to decrease the number of social contacts between people to localize outbreaks and to reduce the possible transmission routes of the pathogen~\cite{flaxman2020estimating,lai2020effect}. Such interventions alone, however, can hardly bring an epidemic to an end, as this requires building high immunity level within the population. This so called herd immunity can be achieved either by the distribution of vaccines, when available, or through the natural spread of the pathogen among the people~\cite{Bjorkman2023,randolph2020herd}. While there is an ongoing debate whether vaccination or natural immunity is more efficient during a pandemic, like the COVID-19 pandemic~\cite{biggs2022vaccination} or HIV~\cite{Steinegger2022}, far less is known about how these immunization strategies perform when they interact with interventions \cite{hale2021global} and the structure of the underlying social network~\cite{gazit2022severe}. We show that in certain cases, natural immunity could lead to better epidemic control of an upcoming outbreak, while vaccine induced immunity can be more effective if it is accompanied with non-pharmaceutical interventions, which effectively decrease social contacts.

We employ a network approach to provide a realistic and quantifiable insight into the outcome of different immunization strategies~\cite{Newman2002,BANSAL2012, Ferrari2006NetworkFrailty, Mann2021SymbioticAntagonistic, Mann2021TwoPathogen, Mann2022NStrain, Hasegawa2011Robustness, Hiraoka2023HerdImmunity}. This approach assumes that the social network of people encodes all possible epidemic routes, highlighting the role of structural~\cite{pastor2015epidemic}, spatial~\cite{balcan2010modeling, Mazzoli2023}, and temporal~\cite{masuda2017temporal} network properties in epidemic spreading. We simulate the epidemic as a spreading process on the network, and we model disease-induced immunity by assuming that infected nodes after recovery become immune to the disease, a process hereafter referred to as \textit{natural immunization}. Vaccination is incorporated as an alternative immunization process, independent of the underlying network structure. Although vaccination behavior is known to depend on various social factors, we restrict our analysis to \textit{random immunization} as a minimal model of vaccine-induced immunity \cite{wang2016statistical}.

However, as found earlier, immunized individuals not only receive direct protection, but they simultaneously act as transmission barriers by blocking and shortening infection pathways~\cite{wang2016statistical,wang2017vaccination}.Thereby, on one hand, they slow down or even prevent the spread of disease to others~\cite{wang2016statistical,mostaghimi2021prevention}; and on the other hand, they create indirect collective protection for the non-immune community, known as the \emph{free rider effect}~\cite{free_rider_1, free_rider_2, free_rider_3}. This effect appears due to the unrolling immunization process, which gradually fragments the underlying network of susceptible individuals, resulting in many disconnected, smaller susceptible components. The magnitude of the free-rider effect, and thereby the effectiveness of the immunization strategy, can be measured by the size of the largest residual susceptible connected component after immunization, also known as the \emph{secondary outbreak size} \cite{Hiraoka2023HerdImmunity, RomeoAznar2022}. Intuitively, effective immunization strategies are able to fragment into small components, with the secondary outbreak size being a worst-case upper bound on the size of a subsequent second wave of the infection, seeded from a single node \cite{Hiraoka2023HerdImmunity}.

Previous work, by Hiraoka et al., compared the random and natural immunization processes and found that they may fragment a social network in markedly different ways, giving rise to two competing network effects \cite{Hiraoka2023HerdImmunity}. In one way, the epidemic tends to infect and thereby immunize highly connected individuals first, effectively turning natural immunization into a form of targeted immunization~\cite{Newman2002,BANSAL2012,Hiraoka2023HerdImmunity}. This mechanism enhances network fragmentation and strengthens disease-induced immunity, especially in networks with greater degree heterogeneity. On the other hand, since natural immunity develops along the pathways of the disease spread, the resulting immune nodes become highly localized, especially in clustered spatial networks. This localization effect weakens the effects of natural immunization, since it reduces the interface between immune and susceptible individuals, as compared to random immunization. As a consequence, the relative performance of the two strategies depends heavily on the heterogeneity of the degree (contact number) distribution and the geographic localization properties of the network. However, this description neglects the effects of non-pharmaceutical interventions, which commonly aim to decrease contact numbers, and this way could crucially change the properties of the underlying network~\cite{perra2021non}. Their effects on the social structure could lead to an overall drop of network density, to increased clustering, to altered degree correlations, and to limited degree heterogeneity, potentially leading to an opposite outcome as compared to earlier predictions. 

In this paper, we study how the effective outcome of simulated random and natural immunization processes is altered when interacting with network interventions. We analyze the stationary state of synthetic epidemic processes on various modeled and real networks, and show that varying network density can change the relative performance of the immunization strategies. Interestingly, we find that in a broad set of networks, uniform vaccination can be more effective than natural immunization, when it is applied simultaneously with non-pharmaceutical interventions.

\section*{Results}

\subsection*{Model design}

We consider a fully susceptible host population whose interactions are encoded by a contact network, with nodes representing individuals and edges representing potential infection transmitting contacts between. To model this structure, we use Geometric Inhomogeneous Random Graphs (GIRGs), a versatile class of geometric scale-free networks with adjustable degree exponent $\gamma$, embedded into the two-dimensional Euclidean space~\cite{GIRG,GIRG2,GIRG3}. The emerging structure of GIRGs is further controlled by a parameter $\alpha \in (1, \infty]$ that scales the influence of spatial proximity on edge formation, potentially leading to locally clustered and globally connected network structures depending on its value. Notably, GIRGs encompass a wide range of well-known network structures as limiting cases; when $\alpha \to 1$ they reduce to non-geometric networks generated by the configuration model, and resemble Random Geometric Graphs~\cite{penrose_rgg} (RGGs) in the joint limit of $\alpha, \gamma \to \infty$ marking an extreme geometric regime. A more detailed description of the GIRG model, along with other network models used in our study is provided in the Methods section. 

In the context of natural immunity, the contact network provides the structure over which the immunity-inducing epidemic process can unfold. We model this process as a susceptible-infected-recovered (SIR) process~\cite{SIR_kermack1927contribution,SIR_anderson1991infectious,pastor2001epidemic_sf}, where each node of the network at a given time can be in one of three mutually exclusive states: susceptible ($\pazocal{S}$), infected ($\pazocal{I}$), or recovered ($\pazocal{R}$). The unfolding of the SIR process is largely determined by a contact-wise transmission probability $\beta$, and a recovery probability $\mu$ (fixed to $\mu=1$) per a unit time (for more precise definition see Methods). Once seeded from a single infected node in an otherwise susceptible population, the infection spreads stochastically through the network until it dies out. Eventually, in our context \emph{natural immunization} against future reinfection arises within the host population for individuals who got infected and subsequently recovered during this first epidemic process, similarly to infectious diseases such as measles, mumps, chickenpox, or rubella~\cite{BANSAL2012}. In contrast, \emph{random immunization} involves selecting individuals (nodes) uniformly at random from the host population and rendering them immune, reflecting the effects of mass vaccination campaigns across a population with no targeted groups.

In our modeling framework, we use the final immunized fraction $f$ as the control parameter instead of fixing the transmission rate $\beta$. We obtain $f$ for natural immunization by simulating spreading processes with $\beta$ sampled uniformly from the unit interval and grouping realizations with the same final fraction $f$. For random immunization, we immunize the same fraction $f$ of uniformly random nodes in the same network. This approach assumes no prior knowledge about $\beta$, which makes the model not only widely applicable, but also mirrors real-world conditions where noisy surveillance data or incomplete reporting often limit accurate estimation of the transmission parameters of a pathogen~\cite{unreliable_param_est_parag2022, unreliable_param_est_SWALLOW2022100547}.

\begin{figure}[h!]
    \centering
\includegraphics[width=1.\linewidth]{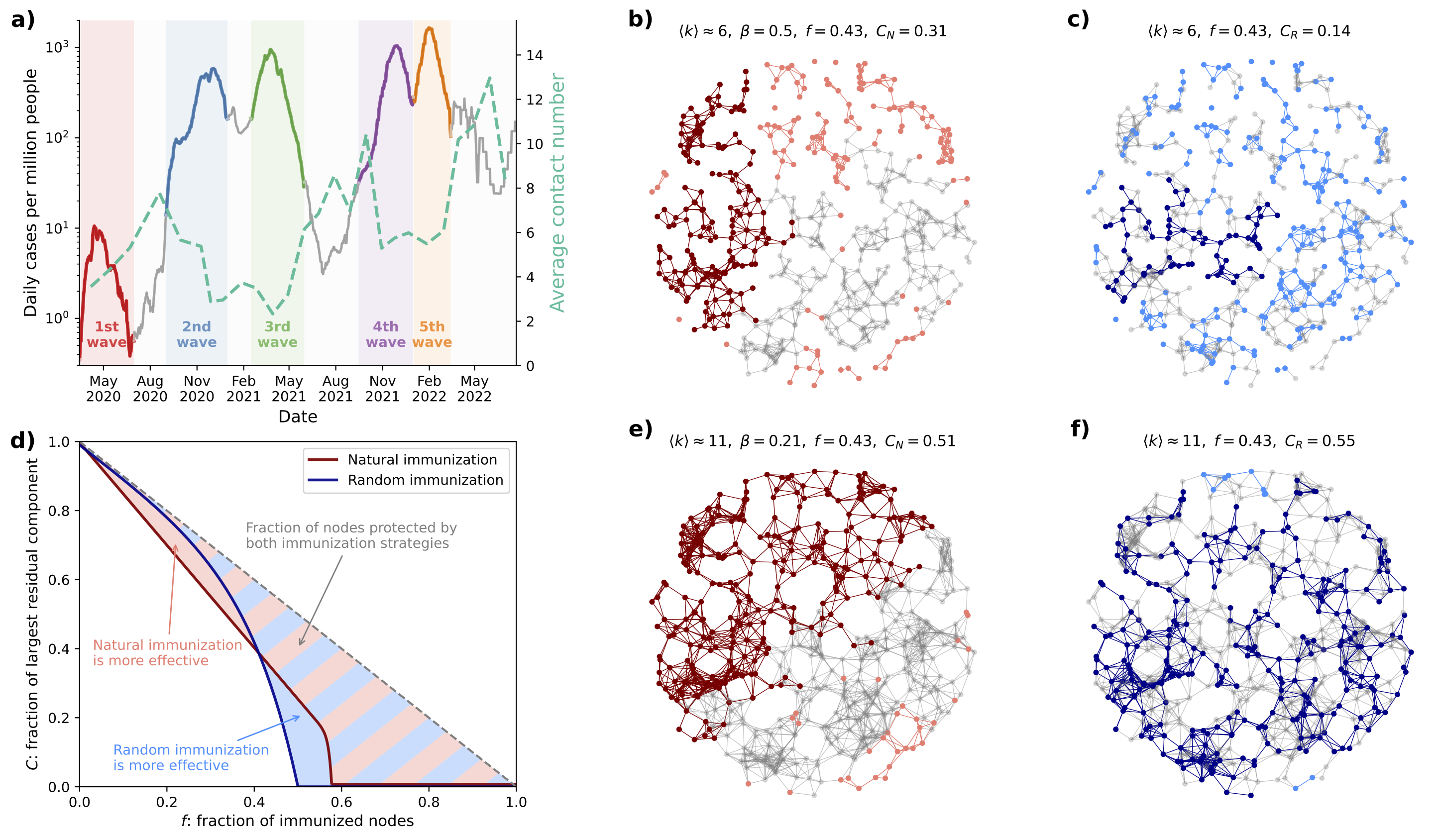}
    \caption{\textbf{Dependence of natural and random immunization effects on network connectivity.} Panel a) shows the daily number of new confirmed COVID-19 cases per million people in Hungary (solid line) together with the average number of contacts of people, as recorded in the MASZK questionnaire (dashed green line), between March 2020 and March 2022. Panels b-c and e-f show the outcomes of natural (left-red) and random (right-blue) immunization on modeled Random Geometric Graphs (RGGs) (modeled as special cases of GIRGs), respectively with $\langle k \rangle = 6$, and $\langle k \rangle = 11$. In all these networks, immune individuals are consistently shown in gray, with their fraction fixed at $f \approx 0.43$. Darker shades highlight the largest residual component, while lighter colors represent the indirectly protected smaller clusters. Panel d) depicts the schematic comparison of natural and random immunization outcomes and the $C$ size of their largest residual component as the function of $f$ fraction of immunized nodes. 
}
    \label{fig:intro_layout}
\end{figure}

\subsection*{Network density impacts immunization efficiency}

Although both the natural and random immunization schemes generate a certain level of protection against an epidemic, their impacts can often be altered by complementary interventions. Non-pharmaceutical interventions typically aim to reduce the number of interactions between individuals, thus suppressing the probability of transmission of the pathogen on the contact network. Such structural changes can be observed in real-world epidemic datasets, as demonstrated in Fig.~\ref{fig:intro_layout}a) through the MASZK COVID-19 longitudinal survey study~\cite{Karsai2020Hungary}. This figure shows a clear anti-correlation between the average number of contacts per person and the incidence of daily new infection cases, a pattern commonly associated with awareness-driven behavioral responses \cite{funk2009spread,odor2025epidemic}. As a result, changes in contact-network density can substantially affect epidemic outcomes and, potentially, the efficacy of immunization strategies.

To quantify the effects of network density on the two immunization strategies, we measure the size of the largest residual components $C_N$ (resp. $C_R$)  remaining after a natural (resp. random) immunization campaign. Since the largest residual component is the maximum size of a second epidemic process (with $\beta=1$), more effective immunization strategies are signified by a smaller residual components $C$. Indeed, a smaller residual component implies a larger fraction of nodes becoming protected by the free rider effect (see Fig.~\ref{fig:intro_layout}d). According to this metric, in the sparser contact networks shown in Fig.\ref{fig:intro_layout}b and c, random immunization is the more effective strategy ($C_N=0.31>0.14=C_R$). This observation aligns with earlier results reported in Ref.~\cite{Hiraoka2023HerdImmunity}. However, this outcome is reversed in the denser networks shown in Fig.\ref{fig:intro_layout}e and f, as the largest residual component is smaller in case of the natural immunization ($C_R=0.55>0.51=C_N$). While these are only example realizations, the observed \textit{strategy reversal} may suggest deeper structural consequences in the immunization patterns and cluster formations due to network density. 

\subsection*{Interplay with network geometry}

\begin{figure}[htb!]
    \centering
\includegraphics[width=1.\linewidth]{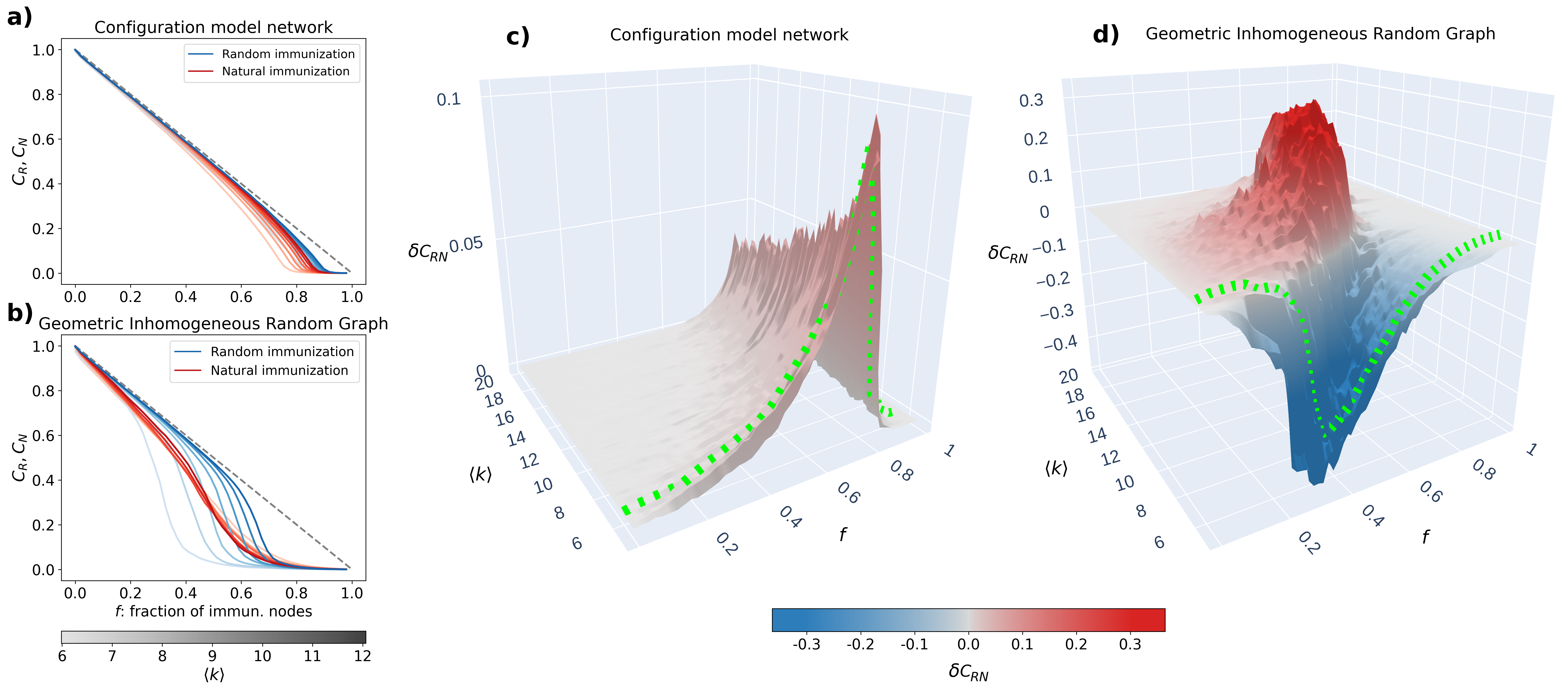}
    \caption{\textbf{Effectiveness of disease- and vaccine-induced immunity across scale-free contact networks with varying average degree $\left<k\right>$ and fixed degree-decay exponent $\gamma = 4$.} Panels a) and b) display the size of the largest residual component under disease-induced (red curves) and vaccine-induced (blue curves) immunization, denoted by $C_N$ and $C_R$, respectively, both plotted as a function of immunization coverage $f$. Results are shown for contact networks with varying average degrees $\left<k\right>$, indicated by different color tones on the gray-scaled colorbar. Panel a) shows the results for non-geometric networks (GIRGs with $\alpha\to 1$), while panel b) displays the corresponding results for their geometric counterparts, generated at $\alpha = 10^6$. Each curve represents averages over five independent network realizations, with 200 SIR processes per network evaluated at 150 transmission probabilities. Panels c) and d) depict the difference in the fraction of indirectly protected individuals between the two immunization strategies, $\delta\pi_{NR}$,  for the same types of contact networks shown as a function of both immunization coverage $f$ and the network’s average degree $\langle k\rangle$. Surface points are colored according to the magnitude and sign of this difference using the blue–red colormap below: red indicates greater indirect protection under natural immunity, whereas blue indicates the opposite. Panel c) showcases results for the configuration-model limit, and panel d) for strongly geometric GIRG networks. Each surface point is obtained from 200 SIR simulations evaluated at 100 evenly spaced transmission probabilities.}
    \label{fig:trend_change}
\end{figure}

Given the strong influence of contact structure on immunization efficiency (Fig.~\ref{fig:intro_layout}), we examine how the interaction of the network density and geometry influence the immunization outcomes. Using synthetically-generated degree-heterogeneous GIRGs, we vary model parameters to assess the effect of network density across different limiting cases of non-geometric and geometric structures. We begin with maximally random networks produced by the configuration model, corresponding to the $\alpha \to 1$ limit of the GIRG model. In this regime, we observe no strategy reversal: natural immunization that performs best remains superior across all densities, regardless of the degree distribution. In Fig.\ref{fig:trend_change}a, the $C_N(f)$ curve shows that the largest residual component size corresponding to natural immunization (in red) consistently lie below the corresponding $C_R(f)$ random immunization curve (in blue) in scale-free networks ($\gamma=4$). In other words, natural immunization consistently outperforms random vaccination for all densities. In Fig.~\ref{fig:trend_change}c, the positive values of the signed difference, $\delta C_{RN}(f)
=C_R(f)-C_N(f)$, representing the difference in the number of indirectly protected nodes, confirm the same result in a three-dimensional visualization. We show that this result remains robust for various degree exponents in the Supplementary Information S3.

In the corresponding geometric scale-free network ($\gamma=4$), generated by the GIRG model in the $\alpha\to\infty$ limit, a more complex trend emerges. As shown in Fig.~\ref{fig:trend_change}b, at low network densities the blue curves fall below the red ones, signaling an advantage of random immunization that persists across the full $f$ range. Nevertheless, this advantage diminishes progressively as the network density increases, indicating natural immunization to become progressively more efficient for denser networks. Eventually, beyond a given average degree, the original trend can even reverse, with natural immunization surpassing random vaccination across the entire range of immunity coverage $f$. The impact of the average degree parameter $\left<k\right>$ on the relative performance of the immunization strategies is even more evident in the three-dimensional plot in Fig.~\ref{fig:trend_change}d. 

Collectively, these findings reveal a notable reversal in immunization efficiency within mildly heterogeneous, yet strongly geometric scale-free networks. Specifically, when the contact network is sparse enough, random vaccination proves to be more efficient in reducing the size of the residual susceptible population at risk of a secondary outbreak. Conversely, as the average contact number increases, disease-induced immunization more effectively limits the susceptible substrate of the population capable of sustaining renewed transmission during subsequent outbreaks. Crucially, this transition is specific to geometric networks and does not occur in non-geometric network models, highlighting the critical role of underlying spatial and structural features in shaping immunization outcomes (see e.g. Fig. S3 in the Supplementary Information).

\subsection*{Structural heterogeneity and surface effects}

From a network-immunization perspective, both disease-induced and random immunization share the same ultimate goal of disrupting large-scale connectivity in the network. However, their efficiency unfolds in different ways depending on the network structure.
As shown earlier in Fig.~\ref{fig:trend_change}c and d, for spatially embedded networks the winning strategy depends on the value of the immunity coverage $f$ and the network density $\left<k\right>$. Therefore, a comparison based on a single value of $f$ could be insufficient and potentially misleading (see Supplementary Information S2). To comprehensively understand the entire trajectory of the immunization dynamics, and to quantify by a single number the overall extent of structural protection offered across all immunity levels, we introduce the aggregate measure
\begin{equation}
\Pi_{j} = \sum_f \pi_j(f) \approx \int_0^1 \pi_j(f)\mathrm{d}f,
\label{eq:cumul_pi}
\end{equation}
which we refer to as the \textit{cumulative structural immunity}. Here, $\pi_j=1 - f - C_j(f)$ denotes the fraction of indirectly protected individuals, with $j \in \{N, R\}$ corresponding to natural and random immunization, respectively. The above quantity is conceptually similar to metrics of network resilience under percolation-based disruptions\cite{ROBUST1, ROBUST2}, highlighting its relevance for structural analyses of epidemic control.

Extending the above approach, we introduce two additional cumulative quantities to uncover the mechanisms behind the observed strategy reversals. First, to quantify the effects of degree heterogeneity, we introduce
$K_{j} = \sum_f \kappa_j(f) \approx \int_0^1 \kappa_j(f)\mathrm{d}f$,
where $\kappa(f)$ is the normalized average degree of immune individuals given by $\kappa_j(f)=\left<k_{j}\right>_{\pazocal{R}}/\left<k\right>$ and $\left<k_{j}\right>_{\pazocal{R}}$ denotes the average degree of recovered $\pazocal{R}$ (immune) nodes in the contact network for $j\in\{N,R\}$. This measure captures the extent to which immunity is biased towards highly connected nodes in the host population. The second measure quantifies the level of geometric localization. Adopting the interface-based perspective introduced in Ref.~\cite{Hiraoka2023HerdImmunity}, we define $P_{j} = \sum_f \rho_j(f) \approx \int_0^1 \rho_j(f)\mathrm{d}f$,
where $\rho_j(f) = E^{(\pazocal{S}\pazocal{R})}_j / E$ measures the fraction of edges connecting susceptible $\pazocal{S}$ and recovered $\pazocal{R}$ (immune) nodes at a given coverage $f$. This measure captures the topological contiguity between the two subpopulations in a cumulative manner, reflecting the degree of mixing at the interface between immunized and susceptible nodes. 

\begin{figure}[htb!]
    \centering
\includegraphics[width=.8\linewidth]{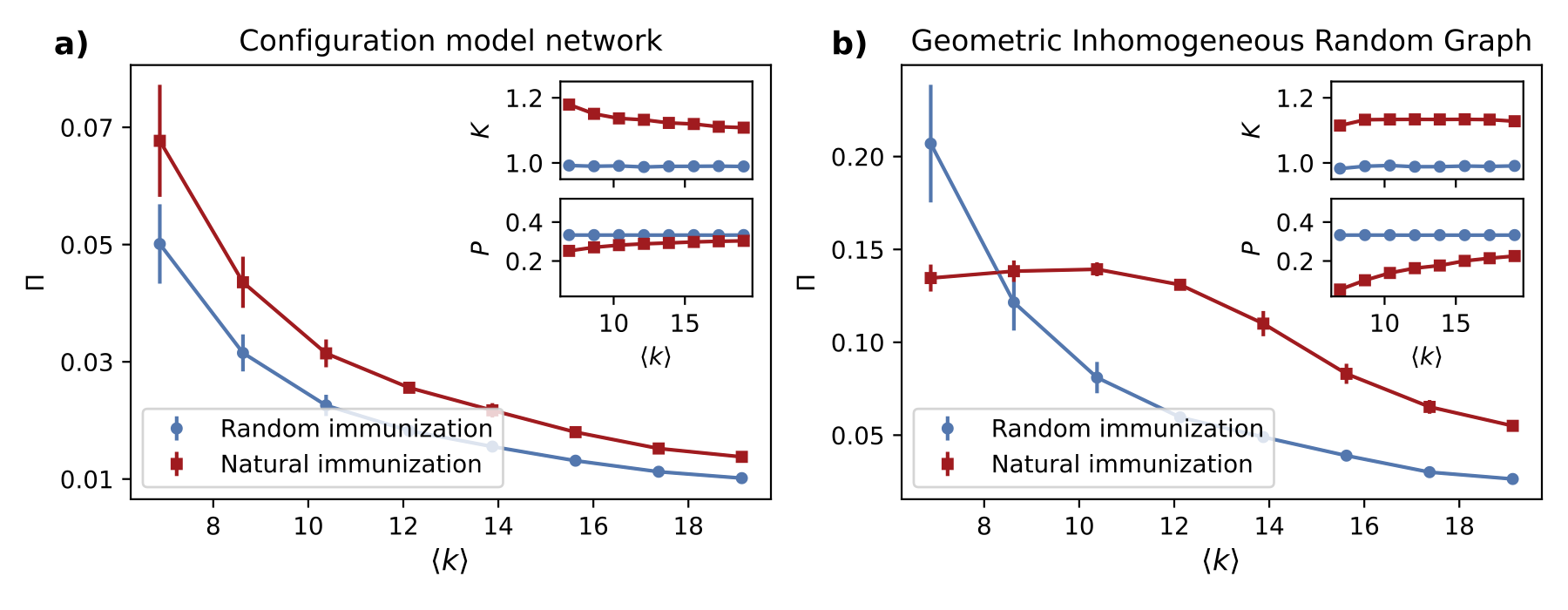}
    \caption{  
    \textbf{Effects of immunization strategies in geometric and non-geometric scale-free networks.} Cumulative structural immunity $\Pi$ (panels a, b), average degree of immunized nodes $K$ (upper insets), and interface area between susceptible and immunized nodes $P$ (lower insets) are shown as functions of the average contact number $\langle k \rangle$ for natural (red) and random (blue) immunity. Panel a) corresponds to strongly geometric GIRGs ($\alpha = 10^6$, $\gamma=4$), while panel b) depicts non-geometric limits ($\alpha \to 1$, $\gamma=4$). The displayed metrics are evaluated across all immunity coverage levels $f \in [0,1]$, and each point shows an average over five independent network realizations. 
    }
    \label{fig:effectiveness_cumuls}
\end{figure}

The measures of $\Pi_{j}$, $K_j$ and $R_j$ altogether offer a complementary, multidimensional view of immunization strategies, jointly capturing structural, degree and localization effects~\cite{Hiraoka2023HerdImmunity} of the underlying network. Their dependence on $\left<k\right>$ in different GIRG models could reveal the origin of the demonstrated strategy reversal for the different immunization strategies. By tracking $\Pi_{j}(\left<k\right>)$, in the configuration model limit (GIRGs with $\alpha \to 1$), where geometric effects are absent, the strategy reversal phenomenon vanishes (see Fig.~\ref{fig:effectiveness_cumuls}a). However, in its geometric counterpart (GIRG with $\alpha\to\infty$, shown in Fig.~\ref{fig:effectiveness_cumuls}b), the reversal phenomenon clearly appears. Following the degree bias function $K_N(\left<k\right>)$ , in the configuration model limit (in Fig.~\ref{fig:effectiveness_cumuls}a upper inset), increasing density progressively reduces the ability of natural immunization to exploit degree heterogeneity, reflected in the systematic decline of $K_N(\langle k\rangle)$. Meanwhile, in geometric networks with the same degree distribution, the degree bias $K_N(\langle k\rangle)$ shows a significantly weaker dependence on density (upper inset of Fig.~\ref{fig:effectiveness_cumuls}b). This behavior resembles lattice-like systems, where there is no preferential immunization due to degree homogeneity, rendering the degree bias completely insensitive to density. This indicates that, in geometric scale-free networks, with $\gamma = 4$, the preferential-immunization mechanism, that would otherwise weaken natural immunization with increasing $\langle k\rangle$, is essentially absent.

On the other hand, while the interface between susceptible and recovered nodes ($P_N$) expands rapidly with $\langle k \rangle$ in geometric graphs (lower inset of Fig.~\ref{fig:effectiveness_cumuls}b), its values remain consistently below those observed in the randomized counterparts (lower inset of Fig.~\ref{fig:effectiveness_cumuls}a). Following Ref.~\cite{Hiraoka2023HerdImmunity}, the rapid increase of the interface $P_N(\langle k\rangle)$ tends to strengthen natural immunization by exposing greater number of paths along which the immunity-inducing epidemic may propagate.

Taken together, the absence of degree bias, and the rapid increase of the susceptible-infected interface leads to a relatively low performance of natural immunization in sparse geometric networks and a relatively high performance in denser ones (Fig.~\ref{fig:effectiveness_cumuls}b). This is in contrast with randomized networks (Fig.~\ref{fig:effectiveness_cumuls}a), where effectiveness decreases steadily with increasing density. The interplay of these mechanisms produces a clear crossover point in Fig.~\ref{fig:effectiveness_cumuls}b, which accounts for the previously observed reversal in the relative performance of the two immunization strategies.

The question remains if the reversal of effective immunization strategies extends to contact networks with different levels of degree heterogeneity, beyond geometric GIRGs with $\gamma = 4$. On one hand, interestingly, we found that less heterogeneous networks, such as Random Geometric Graphs display the same behavior, and even lattices lacking any degree heterogeneity exhibit the observed reversal in the strategies (for results see Supplementary Information S3).

On the other hand, we also examined scale-free networks with stronger degree heterogeneity, i.e. with $3 \le \gamma \le 4$ (see results in Supplementary Figs.~3--4). These simulations reveal two key observations: First, in case of sparse networks with strong geometric coupling ($\alpha = 10^6$) and $\gamma$ close to 4, vaccine-induced immunity outperforms natural immunity, despite the heavy-tailed degree distributions that would lead to an opposite outcome in the non-geometric limit (see the lower curve corresponding to $\langle k \rangle = 6$ in Supplementary Fig.~3c). We note that similar observations have already been reported for spatial networks with negative binomial degree distributions~\cite{Hiraoka2023HerdImmunity}. However, our observations reproduce these results for even higher degree of structural heterogeneity. Second, we observe that natural immunity consistently outperforms random vaccination for denser geometric networks, regardless of the outcome in sparser systems (see the upper curve in Supplementary Fig.~3c). This latter observation again reinforces the general trend that increasing average degree shifts the balance in favor of disease-induced immunization.

\subsection*{Observations in real networks}

While we found the strategy reversal phenomenon to be robust in various synthetic networks, the question remains if they appear in more realistic setups, in empirical networks that are inherently geometric, degree heterogeneous, and exhibits varying link density. While access to such contact networks at scale is not possible, mobility flow networks provide good proxies as they resemble well the spatial embedded social mixing patterns of individuals, and they sensitive to behavioral adaption to non-pharmaceutical interventions. Thus we use mobility networks extracted from the Dynamic Human Mobility Flow Dataset of the United States recorded before and during the COVID-19 pandemic~\cite{US_mobility_flow}. This dataset captures origin-destination flows across the U.S. at multiple spatial resolutions and temporal scales. In our analysis, we utilize weekly aggregated mobility networks at the state level with vertices representing census tracts within the same state and edges being pruned mobility flows between them. Unlike synthetic networks, each instance of these mobility networks represents a unique realization, thus their average degree cannot be adjusted directly. Instead, we exploit that the dataset was recorded between 2019 and 2021 and use it as a natural experiment having the COVID-19 outbreak and the subsequent interventions as treatments. We selected one-week aggregated data starting on 6 April 2019 as a reference period, when there were no travel restrictions and the mobility networks had high link density. We compare these networks with corresponding treatment observations aggregated over one week starting on 8 April 2020, with strict interventions and lockdown measures in place, resulting a significant reduction in the mobility network density. To test whether the strategy reversal phenomenon extends to this empirical data, we simulate epidemic processes on these networks to evaluate the consequences for vaccine- and disease-induced immunization. While mobility networks remain approximations of social mixing patterns, their structure and density (reflecting level of social encounters) is indeed sensitive to non-pharmaceutical interventions. Their adaptive structures strongly influence the outcome of influenza like diseases, making them good candidates to explore the effects of random and natural immunization strategies, altered by different state-level immunization policies applied during a pandemic.

These networks come as almost fully connected and weighted structures with several links characterized by insignificantly small mobility volume. For the purpose of our study, for each state separately, we inferred a specific $w^{*}_{\text{thres}}$ weight threshold that resulted in a sparse but connected mobility network during the pandemic treated period. We used then the same threshold to sparsify the networks during the pre-pandemic period. The steps of data pre-processing and the selection method of the optimal weight threshold are briefly summarized in the Methods section and in the Supplementary Information S4. After simulating epidemic processes on mobility networks, to quantify the differences in the effectiveness of disease- and vaccine-induced immunity, we computed the difference $\Delta \Pi_{NR}=\Pi_{N}-\Pi_{R}$, which is positive if natural immunization is more effective and it becomes negative if random immunization is better.

\begin{figure}[htb!]
    \centering
\includegraphics[width=1.\linewidth]{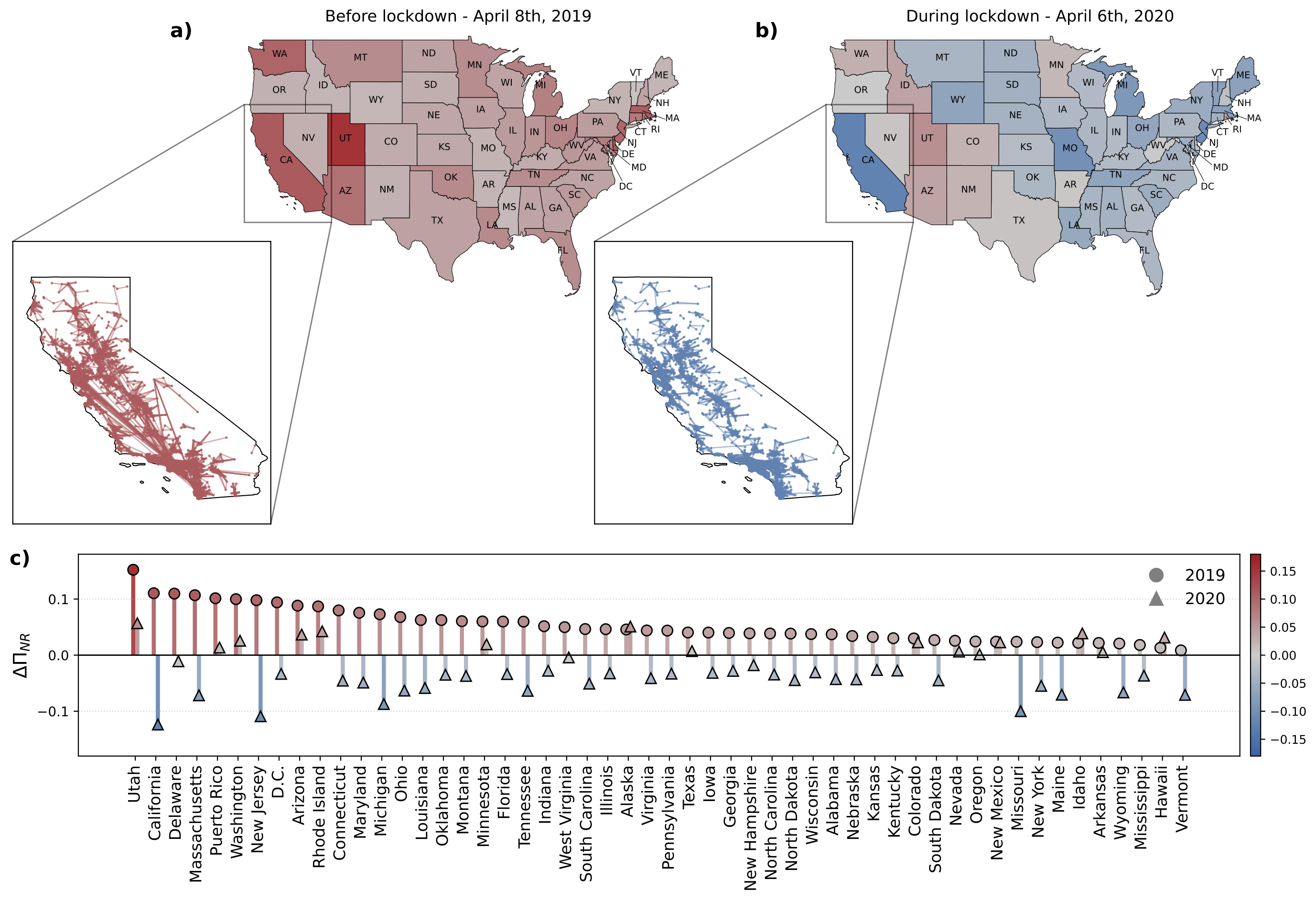}
    \caption{
    \textbf{The relative effectiveness of disease- and vaccine-induced immunization on state-level U.S. mobility networks before and during lockdowns.} Panels a) and b) display the $\Delta\Pi_{NR}$ relative difference in effectiveness values obtained for each state, highlighting the spatial distribution of the effective immunization strategy before and during pandemic interventions. Specifically, panel b) corresponds to the pre-lockdown period in April 2019, while panel b) depicts results during the lockdown period in April 2020. The insets show the layouts of the mobility networks corresponding to California in the two time periods. In all panels, colors indicates the magnitude of $\Delta\Pi_{NR}$, according to the color scale on the right of panel c). Red and blue shows the better effectiveness of natural and vaccine-induced immunity, respectively. Data are displayed for a state-wise optimal threshold value $w^{*}_{\text{thres}}$ (see Methods and Supplementary Information S4). Panel c) shows a bar chart illustrating the differences in effectiveness between natural and vaccine-induced immunity for each U.S. state mobility network with circles representing the pre-lockdown period and triangles indicating the during-lockdown period.
} 
    \label{fig:trend_change_us_mobility}
\end{figure}

Indeed, during the pre-lockdown period we found denser mobility networks in most states as compared to the lockdown period, when strict mobility restrictions were in place. This is demonstrated in Fig.~\ref{fig:trend_change_us_mobility}a) and b) insets, which captures how connections between census tracts in California are effected by interventions; the density reduction was due to several long-range mobility links present in 2019, which were cut off during the 2020 lockdown, rendering the network more geometric and significantly sparser in the latter phase (see Supplementary Information S4 for more details).

The maps shown in Fig.~\ref{fig:trend_change_us_mobility} main panels depict the $\Delta \Pi_{NR}$ values obtained for the pre-pandemic (panel a) and pandemic (panel b) phases in each state. These results suggest that states corresponding to less populated central regions in the U.S. typically exhibit less or not at all the strategy reversal phenomenon, whereas in denser populated states like California or states in the Eastern part of the country, the strategy reversal phenomenon is particularly pronounced in our simulations. This observation suggests that alongside the density reduction, the transition of the network structure from weaker (2019) to stronger (2020) geometric embeddedness can as well contribute to the immunization strategy reversal.

More quantitatively, as shown in Fig.~\ref{fig:trend_change_us_mobility}c, in almost all states the natural immunization (shown as red) became less effective as compared to the pre-pandemic period in 2019 (circles). This leads to the reversal of the effective immunization strategy in several states during the pandemic (triangles). These results overall suggest the same outcomes as our earlier conclusion: vaccination induced immunization (in blue) is more effective when it is applied simultaneously together with interventions reducing network density and amplifying geometric effects. Note that the observations may depend on the distance from the network percolation point, controlled by the mobility weights, as we show in the Supplementary Information S4.

All these results suggest that the strategy reversal phenomenon is not merely a peculiar feature of synthetic geometric networks, but a far more general effect that can arise from complex structural changes in real world networks.

\section*{Discussion}

Non-pharmaceutical interventions aiming to reduce contact rates and limit individual mobility can have significant and even counterintuitive effects on epidemic dynamics. Beyond directly cutting off transmissions, such interventions reshape the map of interactions by directly altering the underlying contact network. A reduction in overall connectivity typically decreases network density, changes local clustering, and other network properties, all of which jointly redefine the pathways through which infections and immunity is distributed. When these structural changes occur in combination with either ongoing or post-immunization processes, their interaction with the applied immunization method can lead to unexpected epidemiological outcomes, diverging from predictions based on static or well-mixed models.

In this work, we explored how different immunization mechanisms respond to structural changes in the host population treated by non-pharmaceutical interventions. We modeled the reduction of contact numbers through two complementary approaches; first, by generating networks with varying density parameters; and second, by analyzing real-world networks observed before and during intervention periods. In both modeling approaches, we found that in geometrically embedded networks, the average contact number plays a crucial role in determining the relative performance of disease-induced versus uniform vaccination strategies. Remarkably, this dependence on network density vanishes in maximally random networks, lacking geometric constraints. This highlights that the interaction of spatial embedding and network density introduces a new additional layer of complexity beyond classical metrics such as degree heterogeneity~\cite{Hiraoka2023HerdImmunity, BANSAL2012, Ferrari2006NetworkFrailty} and clustering~\cite{Mann2021TwoPathogen}.

Our findings extend previous theoretical results by showing that, in geometric networks, the average degree is not only a scaling parameter but a structural determinant that can qualitatively change immunization outcomes. In highly homogeneous systems, e.g. in lattices, natural immunization can be more effective even in the absence of degree heterogeneity, while conversely, in networks exhibiting mild heterogeneity, random vaccination may outperform natural immunization, despite the latter’s preferential targeting of high-degree nodes.

The framework presented here neglects several inherent characteristics of real-world social systems, such as community structure~\cite{girvan2002community} or homophily~\cite{mcpherson2001homophily} in the contact patterns, and the additional costs associated with disease-induced immunity, including morbidity, mortality, and wider economic or societal impacts. Moreover, real epidemic scenarios rarely involve purely natural or vaccine-induced immunity; they rather coexist and interact dynamically within the same host population. Nevertheless, decoupling them and studying their outcomes separately remain still highly relevant. It allows for standardized comparison of their protective effectiveness, helps to clarify under which structural or behavioral conditions one may be more advantageous than the other, and provides helpful insights into the network-level determinants of immune protection.

In summary, our results suggest that interventions altering the average number of contacts -- one of the few network properties that can be controlled to some extent through behavioral changes or policy measures -- can shift the balance between vaccination- and disease-driven immunity. In dense geometric networks, natural immunity ultimately emerges as the more effective strategy, whereas vaccination proves most beneficial when combined with complementary non-pharmaceutical interventions that reduce contact number within the host population. This duality highlights the necessity of integrating network geometry into epidemic control models, particularly when designing synergistic strategies that couple non-pharmaceutical interventions with vaccination campaigns.

\section*{Methods}

\subsection*{Susceptible-Infected-Recovered model on static networks}

We consider the susceptible-infected-recovered (SIR) epidemic process on static contact networks~\cite{SIR_kermack1927contribution,SIR_anderson1991infectious,pastor2001epidemic_sf}. Each node can be in one of three mutually exclusive states: susceptible ($\pazocal{S}$), infected ($\pazocal{I}$), or recovered ($\pazocal{R}$). The dynamics was simulated in discrete time. 
At each time step, every infected node independently transmits the infection to each of its susceptible neighbors with probability $\beta$, after which the newly infected nodes enter the infected state. Simultaneously, infected nodes may recover and transition to the recovered state with probability $\mu$ in one time iteration. Nodes in the recovered state neither acquire nor transmit infection and are thus excluded from further dynamics. The process continues until no infected nodes remain, at which point the system has reached its stationary state. In this study, we consistently set $\mu = 1$.

\subsection*{Simulation of disease-induced immunity}

To characterize disease-induced immunity, we simulate SIR dynamics~\cite{SIR_kermack1927contribution,SIR_anderson1991infectious,pastor2001epidemic_sf} on a given static contact network and let the infection propagate until the system reached a disease free state. In this setup, immune nodes correspond to individuals who got infected and subsequently recovered. The final fraction of immunized individuals $f$ is therefore an emergent quantity determined by the network structure, the stochasticity of the initial infection seed, and the transmission events due to the transmission probability $\beta$. Any quantity of interest $\phi$ is then measured as a function of $f$, that is, $\phi=\phi(f)$. Along this line, given the previous assumptions, two general approaches can be considered; any arbitrary $\phi$ can either be studied as a function of the realized immunity coverage $f$ or indirectly via the transmission probability $\beta$ that generated it. Within the first approach, hereinafter referred to as Method A, $f$ is treated as the main observable control parameter; outcomes from many realizations (corresponding to different seeds and $\beta$ values) are binned by their realized coverage value, and the average $\bar{\phi}(f)$ is computed for each non-empty bin. In contrast, in Method B realizations are instead grouped by the value of $\beta$, and both $\phi$ and $f$ are averaged within each group, yielding $\bar{\phi}(\bar{f}(\beta))$. For a more detailed comparison between Method A and Method B, see the Supplementary Information S1.

In the present study, we adopt Method A, as conditioning on $f$ isolates structural protection at a fixed realized immunity level, independent of the path by which that level was achieved (e.g., different transmission probabilities $\beta$, initial seeds or stochastic histories). This approach avoids mixing up the effects of dynamical parameters and coverage -- important because the mapping between $\beta$ and $f$ can be highly non-linear and noisy especially in strongly geometric networks. Treating $f$ as the control parameter (i) enables fair comparison across immunization mechanisms at equal coverage  and (ii) aligns conceptually better with percolation-style analyses that condition on the realized extent of removal rather than the microscopic rate parameters. In addition, Method A does not rely on any \textit{a priori} knowledge of the epidemic that induces immunity in the contact network, which is particularly relevant in real-world scenarios where information on the transmission dynamics of an ongoing pandemic is unavailable.

\subsection*{Geometric Inhomogeneous Random Graphs (GIRGs)}

The Geometric Inhomogeneous Random Graph (GIRG) model provides a general framework for generating complex networks that display both heterogeneous degree distributions (typically scale-free) and strong geometric embeddedness~\cite{GIRG,GIRG2,GIRG3}. 
This dual structure is motivated by the observation that connectivity in many real-world systems --- ranging from biological to social and technological networks --- depends simultaneously on some intrinsic node properties, such as fitness or popularity, as well as on spatial or similarity-based proximity.

Networks generated by the GIRG model capture this interplay by combining scale-free degree distributions with geometric clustering, thereby offering a parsimonious yet powerful modeling framework for empirical networks. They serve as robust benchmarks for social systems where the likelihood of interaction depends on node proximity in a hidden feature space. Such geometric embeddedness inherently produces a high density of triangles~\cite{clustering_geometry}, strong modularity~\cite{modularity_geometry,modularity_geometry2}, and large graph diameters~\cite{diameter_geometry}; features characteristic of many real-world networks but absent from maximally random graphs generated by the configuration model~\cite{newman2018networks}.

Formally, a GIRG consists of $N$ nodes, each of which is assigned with $d+1$ independent attributes; a weight parameter and a $d$-dimensional position vector in the latent geometric space~\cite{GIRG,GIRG2,GIRG3}. For each node $i$, its weight $w_i \in [w_{\min}, w_{\max}]$ represents the node's intrinsic propensity to establish connections with other nodes. This weight is often assumed to be sampled from a power-law distribution given by
\begin{equation}
    p(w) \sim w^{-\tau}, \quad \tau > 2,
\end{equation}
where the $\tau$ parameter directly controls the heterogeneity of the degree distribution. Nodes with larger weights are significantly more likely to attract connections, yielding to the presence of hubs and producing the characteristic scale-free nature observed in a variety of real-world networks~\cite{barabasi_revmod, barabasi_science}. In addition to weights, node positions are drawn uniformly at random from the $d$ dimensional torus $\mathbb{T}^d = [0,1]^d$ with boundary conditions. This latent embedding defines distances between nodes and captures geometric locality in the generated graphs.

Eventually, the edges between pairs of nodes $(i,j)$ are created independently with a probability that depends on both their intrinsic weights and their geometric proximity. More specifically,
\begin{equation}
    p_{ij} = \min\left\{1, \left( \frac{w_i w_j}{W \cdot \text{dist}_d(x_i, x_j)} \right)^\alpha \right\},
    \label{eq:girg_conn}
\end{equation}
where $\text{dist}_d(x_i, x_j)$ a $d$-dimensional metric on the torus, e.g. the $\infty$-norm or the euclidean distance. $W = \sum_{k=1}^N w_k$ is the sum of weights in the system, and $\alpha > 1$ controls the influence of geometry. Greater values of $\alpha$ strengthen geometric embededness and, hence, local clustering by favoring nearby connections, whereas smaller values induce long-range interactions and reduce spatial constraints in the generated networks.

A notable limiting case of the GIRG model is the Random Geometric Graph~\cite{penrose_rgg} (RGG), which arises when all nodes have identical intrinsic weights ($w_i = w_0$ for all $i$) and the geometric parameter $\alpha$ tends to infinity. In this regime, the connection probability in Eq.~(\ref{eq:girg_conn}) becomes a deterministic function of distance: two nodes are connected if and only if their positions lie within a fixed radius $r$. Consequently, edge formation depends purely on spatial proximity, and the resulting network is entirely determined by positions of the nodes in the underlying geometric space. 

\subsection*{Multiscale Human Mobility Flow dataset of the U.S. during the COVID-19}
\subsubsection*{Data description and preprocessing}
The U.S. mobility data set provides high resolution data on human mobility flows within the United States, covering the time period from January 1, 2019, through the COVID-19 pandemic until April 2021~\cite{US_mobility_flow}. It captures origin-destination flows at three different spatial resolutions, census tract, county, and state levels, which are all inferred by tracking the trajectories of millions of anonymous mobile phone users collected by SafeGraph~\cite{SafeGraph2020}. Mobility flows between geographic units are provided at both daily and weekly time scale at each spatial resolution. Each record in the data contains the geographic identifiers and centroids of the origin and destination units, along with the number of observed visitors and the inferred population flows between them. The dynamic population flows between any pair of origin-destination units, $r_{\text{pop}}(o,d)$, are estimated as
\begin{equation}
r_{\text{pop}}(o,d) = r_{\text{visitors}}(o,d) \times \frac{n_{\text{pop}}(o)}{n_{\text{devices}}(o)}.    
\end{equation}
where $r_{\text{visitors}}(o,d)$ denotes the computed mobile phone-based visitor flow coming from $o$ to $d$, $n_{\text{pop}}(o)$ is the population in the origin unit $o$, and $n_{\text{devices}}(o)$ is the number of unique mobile devices residing in $o$~\cite{US_mobility_flow}. 

For our analysis, we ignore the county level data and focus solely on intra-state mobility dynamics, considering weekly-aggregated flows between census tracts within the same state. We construct a network for each state, where nodes represent census-tracts and edges correspond to mobility flows between them. To study controlled variations in the average degree associated with the observed strategy reversal, we consider two network snapshots per state; a high average degree network from 6 April 2019, well before the COVID-19 outbreak, and a low average degree network from 8 April 2020, during the lockdown period. Nevertheless, in their original form these networks are still weighted and directed~\cite{US_mobility_flow}. To align them with the framework used in this article, we applied a set of pre-processing steps listed below.
\begin{enumerate}
    \item Removing self-connections:
    Self-connections are consistently removed from the networks, considering only mobility flows between different geographical units.
    \item Conversion to undirected networks: 
    Connections are always treated as undirected ones by imposing a fix ordering of origin and destination pairs, hence each link is uniquely represented regardless of its direction.
    \item Aggregation of mobility flows: 
    Links between the same pair of locations are combined, summing their mobility weights to produce a single undirected connection for each pair.
    \item Filtering mobility flows: 
    To convert weighted networks into unweighted ones, only connections with weights above a specified threshold value $w^*_{\text{thresh}}$ are kept, ensuring that the final network contains only significant mobility flows. Once $w^*_{\text{thresh}}$ has been specified, network snapshots from both pre-lockdown (6 April 2019) and during-lockdown (8 April 2020) periods are constructed using this threshold value. For more details about the precise description of this filtering procedure see the following section and the Supplementary Information S4.
\end{enumerate}
Following the pre-processing steps above, we obtain undirected and unweighted networks, where connections represent the most significant intra-state mobility flows between census tracts. The resulting data are then ready for the subsequent network analysis.

\subsubsection*{Filtering mobility flows}

Filtering weights according to their significance allows us to convert the data into undirected networks, which is the input format required for the framework used in this study. Nevertheless, the procedure for finding an optimal threshold is not predefined; to address this, we systematically explored threshold values from a lower bound $w^{\mathrm{min}}_{\mathrm{thres}}=200$ up to an upper bound $w^{\mathrm{max}}_{\mathrm{thres}}$, and performed simulations on the obtained networks for both disease- and vaccine-induced immunization scenarios. 

This yields four distinct immunization curves for each state at any threshold $w\in[w^{\mathrm{min}}_{\mathrm{thres}},\,w^{\mathrm{max}}_{\mathrm{thres}}]$, evaluated with a resolution step of $\delta w=100$. We numerically computed the effectiveness of the two immunization strategies -- vaccine- and disease-induced -- during both in the pre-lockdown (6 April 2019) and during-lockdown (8 April 2020) periods, denoted compactly as $\Pi^{(t)}_{j}$, where $j\in\{N,R\}$ and $t\in\{\text{6 April 2019},\ \text{8 April 2020}\}$.

For each state, the upper bound $w^{\mathrm{max}}_{\mathrm{thres}}$ is chosen such that the largest connected component of the corresponding network remains connected even in the during-lockdown period (8 April 2020), when mobility flows are typically reduced. To allow flexibility in the simulation setup, we defined connectedness of a network, if its largest connected component $S$ is comparable to the total number of nodes $N$, more precisely if $S>0.9N$ holds. The scanned threshold ranges $[w^{\mathrm{min}}_{\mathrm{thres}},\,w^{\mathrm{max}}_{\mathrm{thres}}]$ for each state are summarized in Supplementary Table 1.

Within the studied range of $w$ we define the optimal threshold $w^{*}_{\mathrm{thres}}$ as the weight that produces networks that exhibit the largest change in simulated immunization outcomes between the pre-pandemic and during-pandemic periods. Mathematically, the optimal threshold is given by minimizing the objective
\begin{equation}
w^{*}_{\mathrm{thres}}
\;=\;
\arg\min_{w}\left\{ 
\omega_{\mathrm{flip}}\times 
\frac{\Delta\Pi^{(2020)}_{NR}-\Delta\Pi^{(2019)}_{NR}}
{\bigl|\Delta\Pi^{(2020)}_{NR}+\Delta\Pi^{(2019)}_{NR}\bigr|}
\right\},
\label{eq:mobility_tch_metric}
\end{equation}
where $\Delta\Pi^{(t)}_{NR}$ denotes the signed difference in effectiveness between the two immunization strategies at date $t$ with $t\in\{\text{6 April 2019},\ \text{8 April 2020}\}$. Positive and negative values of $\Delta\Pi^{(t)}_{NR}$ indicate the superiority of disease-induced and vaccine-induced immunization, respectively. The quantity of $\omega_{\mathrm{flip}}$ appearing in Eq.(\ref{eq:mobility_tch_metric}) is a weighting factor defined as
\begin{equation}
    \omega_{\text{flip}} =
\begin{cases}
1, & \text{if } \operatorname{sgn}(\Delta \Pi^{(2020)}_{NR}) \neq \operatorname{sgn}(\Delta \Pi^{(2019)}_{NR}), \\[0.5em]
0, & \text{if } \operatorname{sgn}(\Delta \Pi^{(2020)}_{NR}) = \operatorname{sgn}(\Delta \Pi^{(2019)}_{NR}),
\end{cases}
\end{equation}
used to penalize $w$ values, where there is no observable strategy reversal between the pre- and during-lockdown periods, i.e., where the same immunization scheme wins in both periods. 

\bibliography{references_.bib}

\clearpage
\thispagestyle{empty}  

\vspace*{0.5cm}  

\begin{center}
{\LARGE\bf Non-Pharmaceutical Interventions Reshape Network Immunization Outcomes -- Supplementary Information\par}
\vspace{0.5cm}

{\large 
Sámuel G. Balogh$^{1,2}$, 
Gergely Ódor$^{1,3}$, 
Márton Karsai$^{1,4,*}$ 
\par}
\vspace{0.5cm}

{\it
$^1$National Laboratory of Health Security, HUN-REN Alfréd Rényi Institute of Mathematics, Budapest, 1053, Hungary\\
$^2$Faculty of Electrical Engineering, Mathematics and Computer Science, Delft University of Technology, 2600 GA Delft, The Netherlands\\
$^3$Institute for Hygiene and Applied Immunology, Medical University of Vienna, Vienna, 1090, Austria\\
$^4$Department of Network and Data Science, Central European University, Vienna, 1100, Austria
\par}
\end{center}
\vspace{1.cm}

\renewcommand{\thefigure}{S\arabic{figure}}
\setcounter{figure}{0}
\renewcommand{\thetable}{S\arabic{table}}
\setcounter{table}{0}
\renewcommand{\theequation}{S\arabic{equation}}
\renewcommand{\thesection}{S\arabic{section}}
\setcounter{equation}{0}
\renewcommand{\thesubsection}{s\arabic{subsection}}
\renewcommand{\thesubsubsection}

\section{Averaging methods in natural immunity simulations}

As outlined in the Methods section, any arbitrary property $\phi$ of the immunization strategies can either be studied as a function of the realized immunity coverage $f$ (Method A) or indirectly via the transmission probability $\beta$ that generated it (Method B). In contrast to Ref.~\cite{Hiraoka2023HerdImmunity}, which consistently employs Method B, the present study adopts Method A as the general averaging approach. In this setup, natural (disease-induced) immunity is simulated as follows; assuming the contact network is given, we first sample transmission probabilities
\begin{equation}
\beta_i = \frac{i}{n_\beta}, \quad i = 1, \dots, n_\beta,
\label{eq:transmiss}
\end{equation}
and for each $\beta_i$, we perform $n_S$ independent simulations with different epidemic seeds and stochastic histories, indexed by $j = 1, \dots, n_S$, yielding
\begin{equation}
\beta_i^j, \quad i = 1, \dots, n_\beta, \; j = 1, \dots, n_S.
\end{equation}
For each sampled $\beta_i^j$, an SIR model is simulated resulting in an immunity coverage $f_i^j \in (0,1]$ that depends on the transmission probability, the initial seed and the stochastic evolution of the epidemic. The admissible range of coverage values is then discretized into bins, and any quantity of interest $\phi$ (e.g., $C_N$, $\kappa_N$, or $\rho_N$, as defined in the main text) is averaged over all simulations $f_l^k$ that fall within a given bin $[f^*, f^* + \delta f^*)$, i.e.,
\begin{equation}
\phi(f^*) = \frac{1}{\left| \{ (k,l) : f_l^k \in [f^*, f^* + \delta f^*) \} \right|}
\sum_{(k,l) : f_l^k \in [f^*, f^* + \delta f^*)} \phi(f_l^k).
\end{equation}
In contrast, Method B, as used in Ref.~\cite{Hiraoka2023HerdImmunity}, follows a different approach. It averages over transmission probability values $\beta$. More specifically, the values of $\beta_i$ are sampled as in Eq.~(\ref{eq:transmiss}), but the resulting immunity coverage $f_l^{*}$ is obtained as the mean of all realizations $f_i^{j}$ generated under the same transmission probability value,
\begin{equation}
    f_l^{*} = \frac{1}{n_S} \sum_{j=1}^{n_S} f_l^{j}(\beta^{j}_l),
\end{equation}
and any derived quantity $\phi$ is then expressed as a function of this averaged coverage, $\phi(f_l^{*})$. 

\begin{figure}[htb!]
    \centering
\includegraphics[width=1.\linewidth]{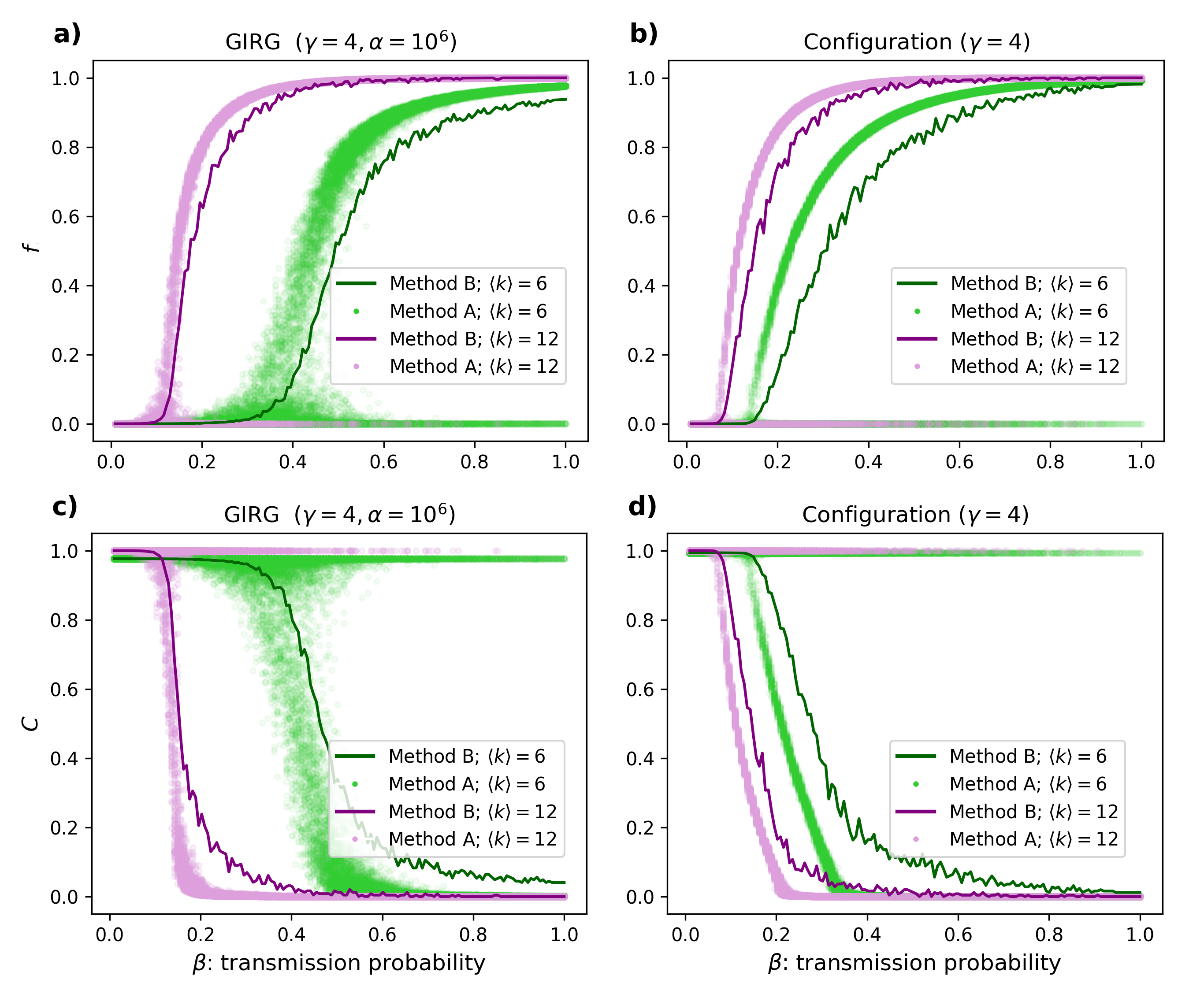}
    \caption{\textbf{Visualization of averaging methods and the relationship between transmission probability and the final fraction of immunized individuals under natural immunization.}  Panels a), b) depict the relationship between the transmission probability $T$ of the immunity-inducing epidemics and the stationary fraction of immunized individuals $f$.  Panels a), c) shows results for a strongly geometric GIRG ($\alpha = 10^6$), while panels b), d) corresponds to the configuration model limit of GIRGs ($\alpha = 1$).  Networks have size $N = 10^4$ and degree exponent $\gamma = 4$. Scatter points show the averages computed using Method A, whereas solid lines represent the outcomes of Method B, highlighting how the choice of averaging procedure affects the binning of the order parameter $f$. Green color correspond to sparser networks ($\langle k \rangle = 6$) while purple color to denser networks ($\langle k \rangle = 12$), illustrating the influence of contact density on the final fraction of immunized individuals.
    }
    \label{fig:avg_method}
\end{figure}

A visual comparison of the two averaging approaches is displayed in Fig.~\ref{fig:avg_method}. This figure compares the key differences between the two methods: Method B establishes a one-to-one correspondence between $\beta$ and $f$, where each value of $f$ is associated with a unique $\beta$. Nevertheless, Method A allows a given $f$ to emerge from a wide range of $\beta$ values (see Fig.~\ref{fig:avg_method}~a--b). Moreover, as shown in panels a--d of Fig.~\ref{fig:avg_method}, in non-geometric networks a given $\beta$ leads to less variation in the resulting $f$ and $C$ values (panels a, c), while in geometric networks the data points are more widely scattered around their mean (panels b, d). The latter observation reflects the presence of bottlenecks in geometric networks, where the immunity-inducing epidemic can easily become trapped ultimately leading to smaller values of $f$.

\section{Global measures to evaluate efficiency of immunization protocols}
An in-depth comparison of disease- and vaccine-induced immunization strategies requires analyzing the entire functional relationship of the respective $\pi_j(f)$ curves for $j\in\{N,R\}$, rather than drawing conclusions based on a single $f$ point. To eliminate this dependence and quantify the overall effect of herd immunity with a single scalar metric, perhaps the most straightforward and commonly used approach is to employ the herd immunity threshold~\cite{Hiraoka2023HerdImmunity,fine2011herd}, defined as
\begin{equation}
f_c = \min \{ f \mid C(f) = 0 \}.
\label{eq:HIT}
\end{equation}
This threshold represents the minimum level of immunization coverage required to completely fragment the network, thereby preventing any future macroscopic-scale outbreak. Nevertheless, this quantity has important limitations. First, not all epidemics are supercritical; depending on the model parameters or various real-world conditions, the epidemic that induces natural immunity in a population may result in substantially different final prevalence levels, and thus very different values of $f$. This indicates that coverage values of $f$ that are far below the herd immunity threshold $f_c$ should also be taken into account when comparing different immunization schemes, an aspect that cannot be captured by Eq.(\ref{eq:HIT}). Additionally, the second limitation of the structural threshold $f_c$ is that it neglects situations where the largest residual component becomes very small, but remains non-zero, i.e., when the network is nearly, but not entirely fragmented by immunization. Furthermore, in numerical simulations, the determination of $f_c$ is sensitive to the choice of a small numerical value $\epsilon$ used to approximate zero, i.e.,
\begin{equation}
f_c(\epsilon) = \min \{ f \mid C(f) < \epsilon \}.
\end{equation}
This issue is particularly relevant in cases where immunization curves lack a sharp critical point, but instead exhibit a gradual transition~\cite{percolation_self_sim}, reminiscent of smeared phase transitions~\cite{smeared_phase}. It is also important in relatively small systems, where finite-size effects play a significant role, as is the case for the mobility networks of many U.S. states in the SafeGraph dataset~\cite{SafeGraph2020}. In light of these limitations, we omit the use of the hard structural herd immunity threshold defined by Eq. (\ref{eq:HIT}) and capture the effectiveness of immunization schemes in a broader context. A simple cumulative metric that does not suffer from the above drawbacks and represents the overall effectiveness of an immunization scheme across the whole range of immunization coverage can be defined by Eq.(1) in the main text. The integral in Eq.(1) quantifies the average fraction of indirectly protected individuals across all immunity levels, providing a robust, single-value summary of the immunization scheme's effectiveness. 

\section{Additional simulation results on synthetic networks}

In this section, we present simulation results that complement the analysis in the main text. Specifically, we investigate the effectiveness of natural and random immunization strategies beyond scale-free (SF) networks with $\gamma = 4$. To this end, we mainly perform side-by-side comparisons of the sizes of the largest residual components, $C_i$ with $i \in \{N, R\}$, along with the signed difference in the number of indirectly protected individuals, defined as
\begin{equation}
\delta \pi_{NR} = \pi_N - \pi_R
\end{equation}
where $\pi_i(f) = 1 - f - C_i(f), \; i \in \{N, R\}$ in agreement with Eq.(1) in the main text. In addition to these point-wise comparisons, we also consider a cumulative difference measure, defined as
\begin{equation}
    \Delta \Pi_{NR} = \Pi_N-\Pi_R = \int_{0}^{1} (\pi_N(f) - \pi_R(f)) \, \mathrm{d}f,
\end{equation}
to quantify the overall gap between the effectiveness of the strategies across a wider range of immunity coverage. 

We begin with the analysis of degree-homogeneous and moderately heterogeneous networks and then turn to the results obtained for SF networks with different degree-decay exponents $\gamma$. 

\subsection{Periodic lattices and random regular graphs}

\begin{figure}[h!]
    \centering
\includegraphics[width=.75\linewidth]{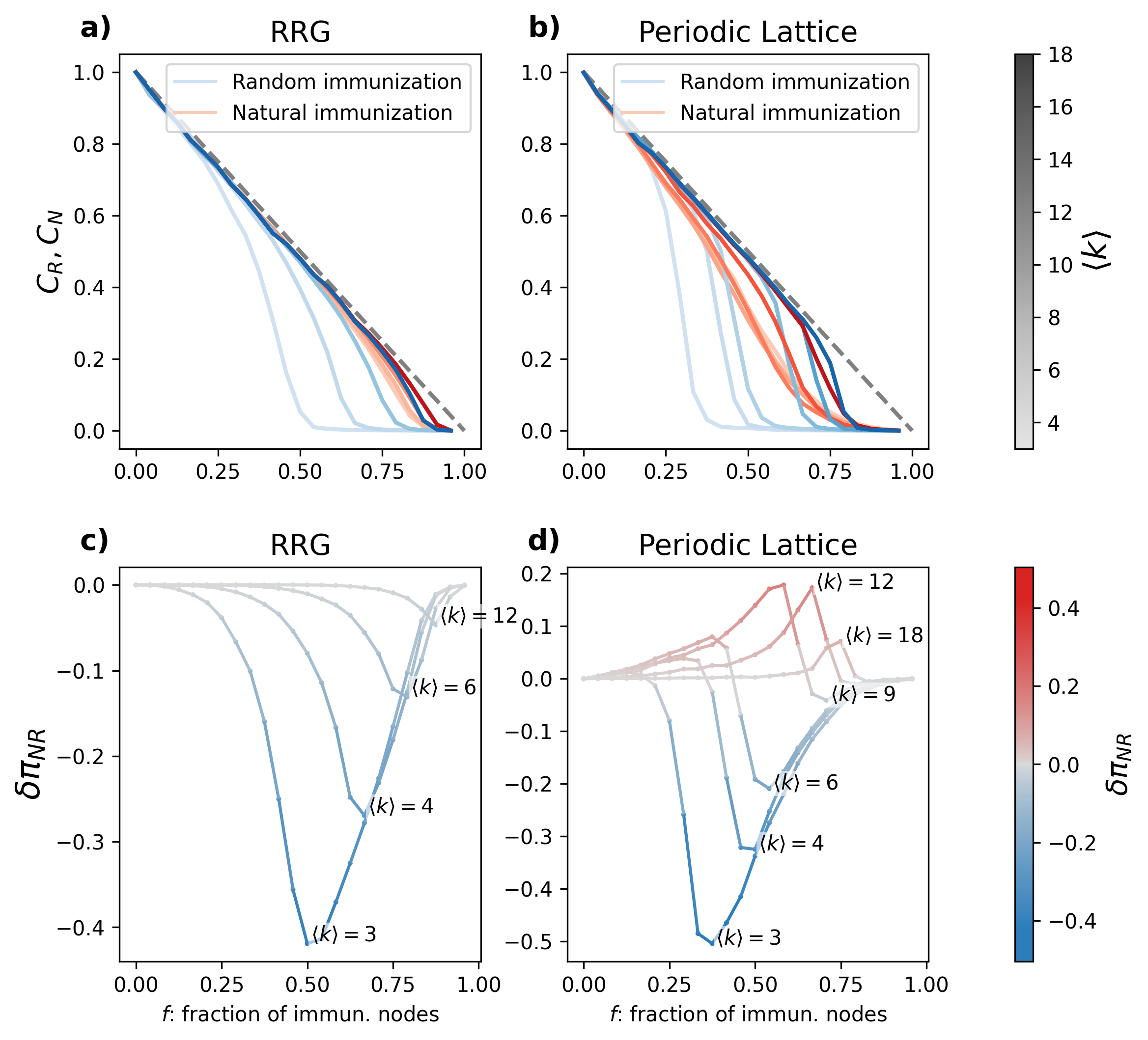}
    \caption{\textbf{Effectiveness of natural and random immunization across degree-homogeneous contact networks with varying average degree $\left<k\right>$.} Panels a) and b) display the size of the largest residual component under natural (red curves) and random (blue curves) immunization, denoted by $C_N$ and $C_R$, respectively, both plotted as a function of immunization coverage $f$. Results are shown for contact networks with varying average degrees $\left<k\right>$, indicated by different color tones on the upper grey colorbar. Panel a) show the results random regular graphs (RRG), while panel b) displays the corresponding results for their geometric counterparts, i.e. 2 dimensional periodic lattices. Panels c) and d) quantify the relative difference in the fraction of indirectly protected individuals, measured as $\delta\pi_{NR}=\pi_N - \pi_R$, across the same set of average degree values. Here, the curves are color-mapped according to the magnitude and sign of this difference, as shown by the lower blue-red colorbar. Red color corresponds to a higher level of indirect protection under natural immunity, while blue indicates the opposite. Similarly to the upper panels, panel c) corresponds to RRGs and panel d) to periodic lattices. Each curve in a)--d) shows simulation results averaged over five independent network realizations, each containing 200 differently seeded SIR processes evaluated at 150 different transmission probabilities.}
    \label{fig:trend_change_lattice}
\end{figure}

First, let us examine two dimensional lattices and their randomized counterparts random regular graphs both corresponding to the degree-homogeneous case. To study how natural and random immunization depend on the average degree in lattices, we require lattice graphs whose densities can be tuned systematically. Probably the simplest and geometrically most faithful way to achieve this is to progressively connect higher order neighbors, i.e. nodes separated by more than one lattice spacing. In a two-dimensional lattice, each node connects with links only to its nearest neighbors. Systematic densification proceeds by adding further edges to next-nearest neighbors, third-nearest neighbors, and so forth, following the inherent geometric ordering of inter-node distances. Each successive order of connectivity increases the average degree while preserving the underlying lattice symmetry, allowing the density of the graph to be tuned in a controlled manner. This augmentation of connections provides a simple yet flexible mechanism for interpolating between sparse regular structures and increasingly dense lattice-like graphs, enabling systematic exploration of how immunization patterns evolve under controlled increases in local connectivity. To generate lattices with average degrees $\langle k\rangle \in \{3, 4, 6, 9, 12, 18\}$, we constructed the following regular two-dimensional lattices; \textit{hexagonal (honeycomb) lattice}, which yields an average degree of $3$, \textit{square lattice}, producing an average degree of $4$, and a \textit{triangular lattice}, in which each node has $6$ neighbors. To obtain denser lattice variants with $\langle k\rangle = 9, 12,$ and $18$, we augmented each of the above lattices by connecting nodes to their \textit{second-nearest neighbors}, that is, by constructing the second power of each lattice. All lattice constructions were implemented with periodic boundary conditions to reduce finite-size effects. The numerical results for the effectiveness of natural versus random immunization as a function of the average degree are displayed in Fig.~\ref{fig:trend_change_lattice}b,d, with panels b and d showing the size of the largest residual component and the number of indirectly protected individuals, respectively. As can be observed in these figures, random immunization performs better at low average degrees, whereas lattice densification causes natural immunization to become more effective. In other words, similar to strongly geometric GIRGs ($\alpha = 10^6$) with a degree exponent $\gamma=4$, regular lattices also exhibit the strategy reversal phenomenon. In contrast, the randomly rewired counterparts corresponding to random regular graphs do not exhibit such a reversal, as shown in Fig.~\ref{fig:trend_change_lattice}a,c.

\subsection{Random Geometric Graphs and Erdős-Rényi graphs}

\begin{figure}[h!]
    \centering
\includegraphics[width=.75\linewidth]{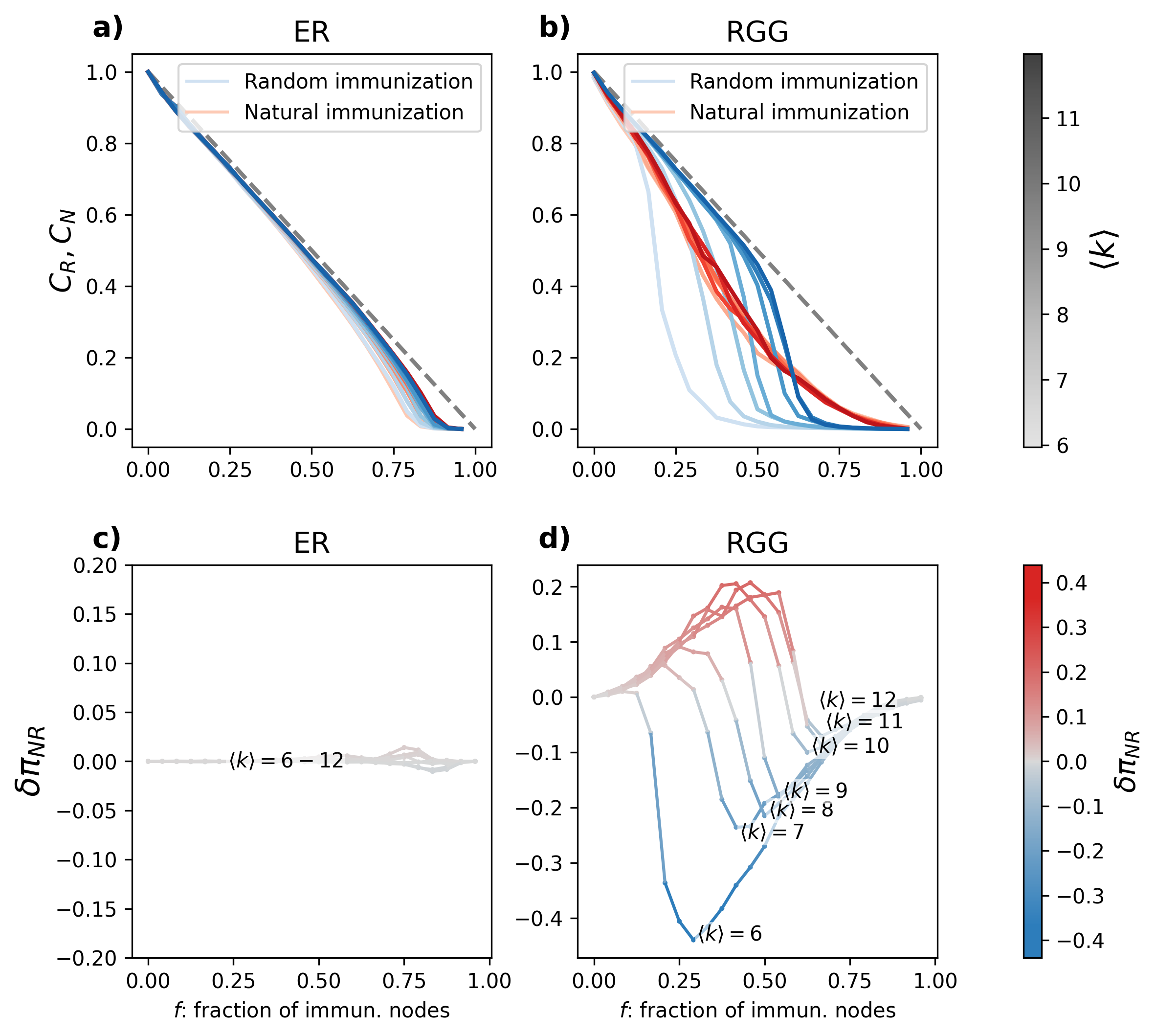}
    \caption{\textbf{Effectiveness of natural and random immunization in Erdős-Rényi and Random geometric graphs with varying average degree $\left<k\right>$.} Panels a) and b) display the size of the largest residual component under natural (red curves) and random (blue curves) immunization, denoted by $C_N$ and $C_R$, respectively, both plotted as a function of immunization coverage $f$. Results are shown for contact networks with varying average degrees $\left<k\right>$, indicated by different color tones on the upper grey colorbar. Panel a) shows the results for random regular graphs (RRG), while panel b) displays the corresponding results for their geometric counterparts, i.e. 2 dimensional periodic lattices. Panels c) and d) quantify the relative difference in the fraction of indirectly protected individuals, measured as $\delta\pi_{NR}=\pi_N - \pi_R$, across the same set of average degree values. Here, the curves are color-mapped according to the magnitude and sign of this difference, as shown by the lower blue-red colorbar. Red color corresponds to a higher level of indirect protection under natural immunity, while blue indicates the opposite. Similarly to the upper panels, panel c) corresponds to RRGs and panel d) to periodic lattices. Each curve in a)--d) shows simulation results averaged over five independent network realizations, each containing 200 differently seeded SIR processes evaluated at 150 different transmission probabilities.}
    \label{fig:trend_change_rgg}
\end{figure}

We next move from maximally homogeneous networks to a mildly heterogeneous regime, considering Random Geometric Graphs (RGG) and Erdős–Rényi random graphs (ER). These networks are still largely homogeneous, although their Poisson-like degree distribution introduces moderate degree variability. The resulting dependence of the strength of natural and random immunization on the average degree is shown in Fig.~\ref{fig:trend_change_rgg}. Similarly to maximally degree-homogeneous structures (lattices, random regular graphs), the strategy reversal phenomenon is clearly observed in the geometric case (RGG), whereas no such reversal occurs in non-geometric ER networks.

\subsection{Geometric Inhomogeneous Random Graphs}

\begin{figure}[h!]
    \centering
\includegraphics[width=.75\linewidth]{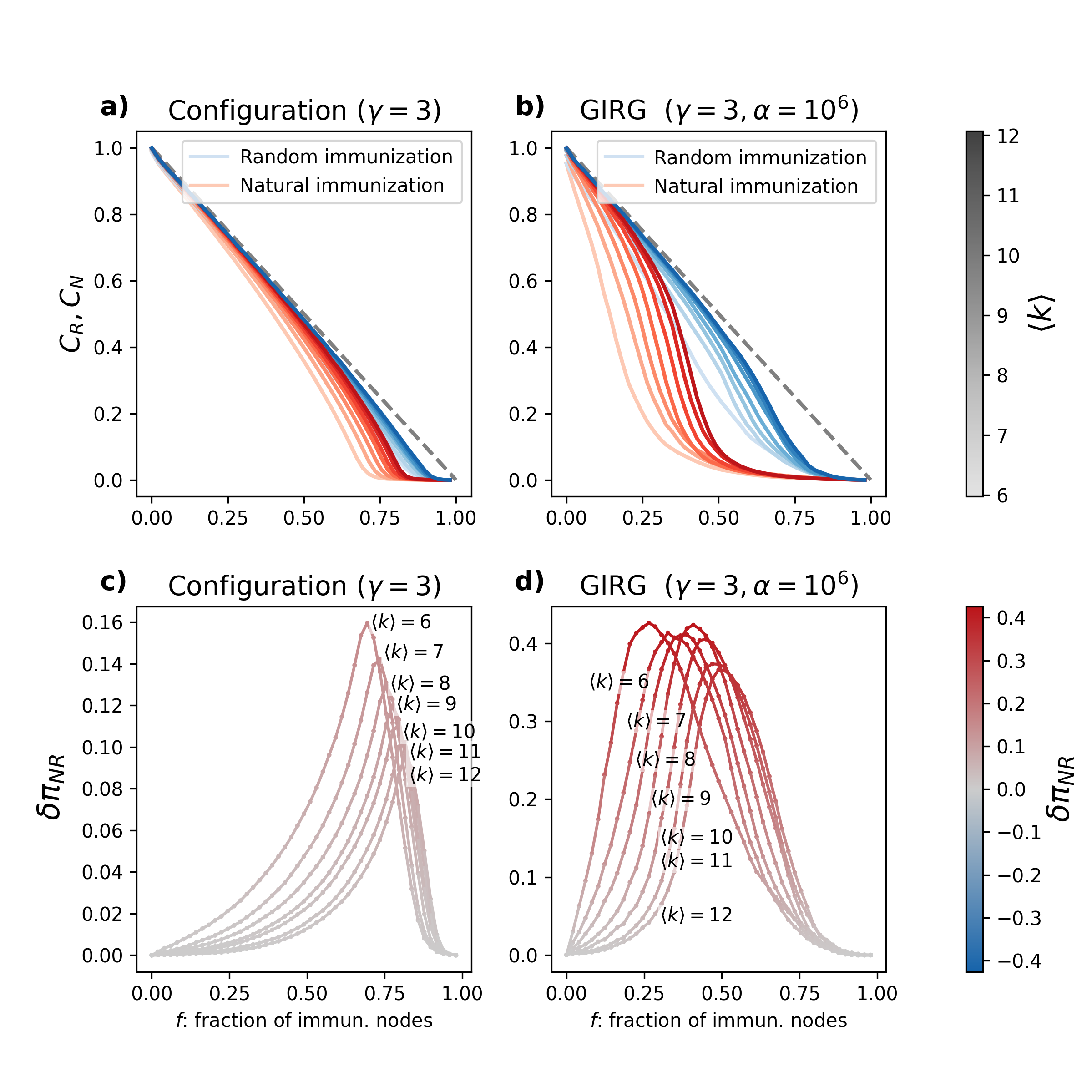}
    \caption{\textbf{Effectiveness of natural and random immunization across scale-free contact networks with varying average degree $\left<k\right>$ and fixed degree-decay exponent $\gamma = 3$.} Panels a) and b) display the size of the largest residual component under natural (red curves) and random (blue curves) immunization, denoted by $C_N$ and $C_R$, respectively, both plotted as a function of immunization coverage $f$. Results are shown for contact networks with varying average degrees $\left<k\right>$, indicated by different color tones on the upper grey colorbar. Panel a) shows the results for networks constructed via the configuration model, while panel b) displays the corresponding results for their geometric counterparts, generated by the Geometric Inhomogeneous Random Graph (GIRG) model. Panels c) and d) quantify the relative difference in the fraction of indirectly protected individuals, measured as $\delta\pi_{NR}=\pi_N - \pi_R$, across the same set of average degree values. Here, the curves are color-mapped according to the magnitude and sign of this difference, as shown by the lower blue-red colorbar. Red color corresponds to a higher level of indirect protection under natural immunity, while blue indicates the opposite. Similarly to the upper panels, panel c) corresponds to configuration model networks and panel d) to strongly geometric GIRG networks. Each curve in a)--d) shows simulation results averaged over five independent network realizations, each containing 200 differently seeded SIR processes evaluated at 150 different transmission probabilities.}
    \label{fig:trend_change_g3}
\end{figure}

Heterogeneous contact networks with power-law degree-distributions yield more diverse outcomes. In a somewhat similar fashion to Fig.~2a-d in the main text -- which shows results for scale-free networks characterized by a degree-decay exponent of $\gamma = 4$ -- we present here an extended set of simulations for both geometric and non-geometric GIRGs~\cite{GIRG,GIRG2,GIRG3} with degree exponent $\gamma = 3$. Figure~\ref{fig:trend_change_g3} reports the size of the largest residual component under natural and random immunization (panels a–b), as well as the corresponding differences in the fraction of indirectly protected, or free-rider individuals (panels c–d). As in the case of GIRGs with $\gamma = 4$, the non-geometric limit (configuration-model) shows no observable evidence of a reversal in the relative effectiveness of the two immunization strategies when $\gamma = 3$ (see Fig.~\ref{fig:trend_change_g3} panels a, c). However, in particular, and in contrast to the results obtained for $\gamma = 4$, the geometric GIRGs with $\gamma = 3$ also do not display any trend reversal (see Fig.~\ref{fig:trend_change_g3} b, d), despite the fact that such a phenomenon is present in the geometric regime for $\gamma = 4$.

\begin{figure}[h!]
    \centering
\includegraphics[width=.75\linewidth]{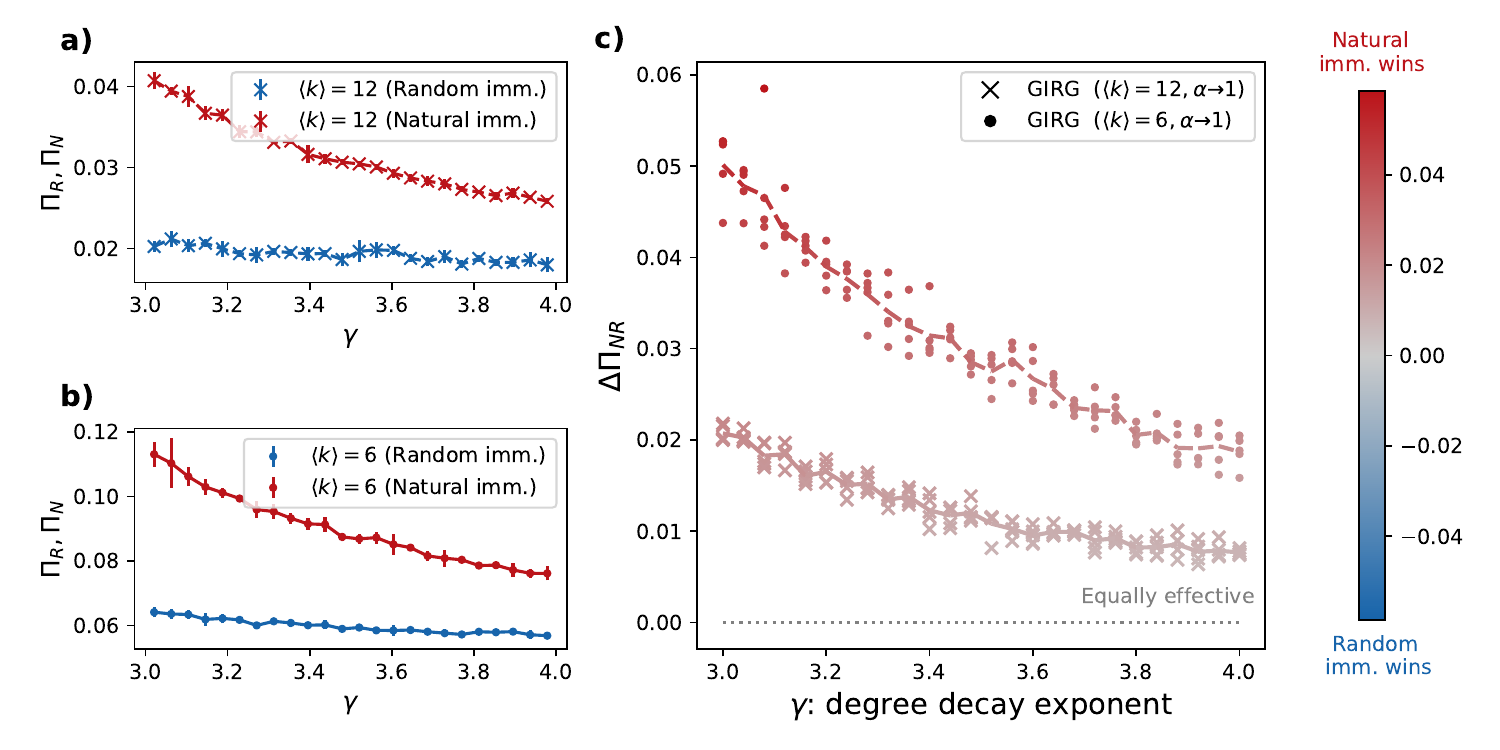}
    \caption{\textbf{Cumulative herd immunity from naturally and randomly acquired protection in the non-geometric limit of GIRG networks.} Panels a) and b) show cumulative herd immunity from natural ($\Pi_N$, red) and random ($\Pi_R$, blue) immunization as a function of the degree-decay exponent $\gamma$ in dense ($\langle k \rangle = 12$) and sparse ($\langle k \rangle = 6$) contact networks, respectively. Networks are generated using the Geometric Inhomogeneous Random Graph (GIRG) model with $N= 10^4$ and $\alpha \to 1$. Panel c) displays the difference between natural and random herd immunity levels (captured by $\Delta\Pi_{NR}=\Pi_N - \Pi_R$) for both the denser and sparser case, with values color-coded according to the colorbar on the right side of panel c). In panels a) and b) each point corresponds to an average over five independent network realizations. In panel c) results for individual networks are visualized as a scatter plot (points and crosses for sparse and dense networks, respectively), while color-coded lines indicate their averages. The grey dotted line denotes equality between the two strategies $\Delta\Pi_{NR}=0$.}
    \label{fig:effectiveness_2}
\end{figure}

\begin{figure}[h!]
    \centering
\includegraphics[width=.75\linewidth]{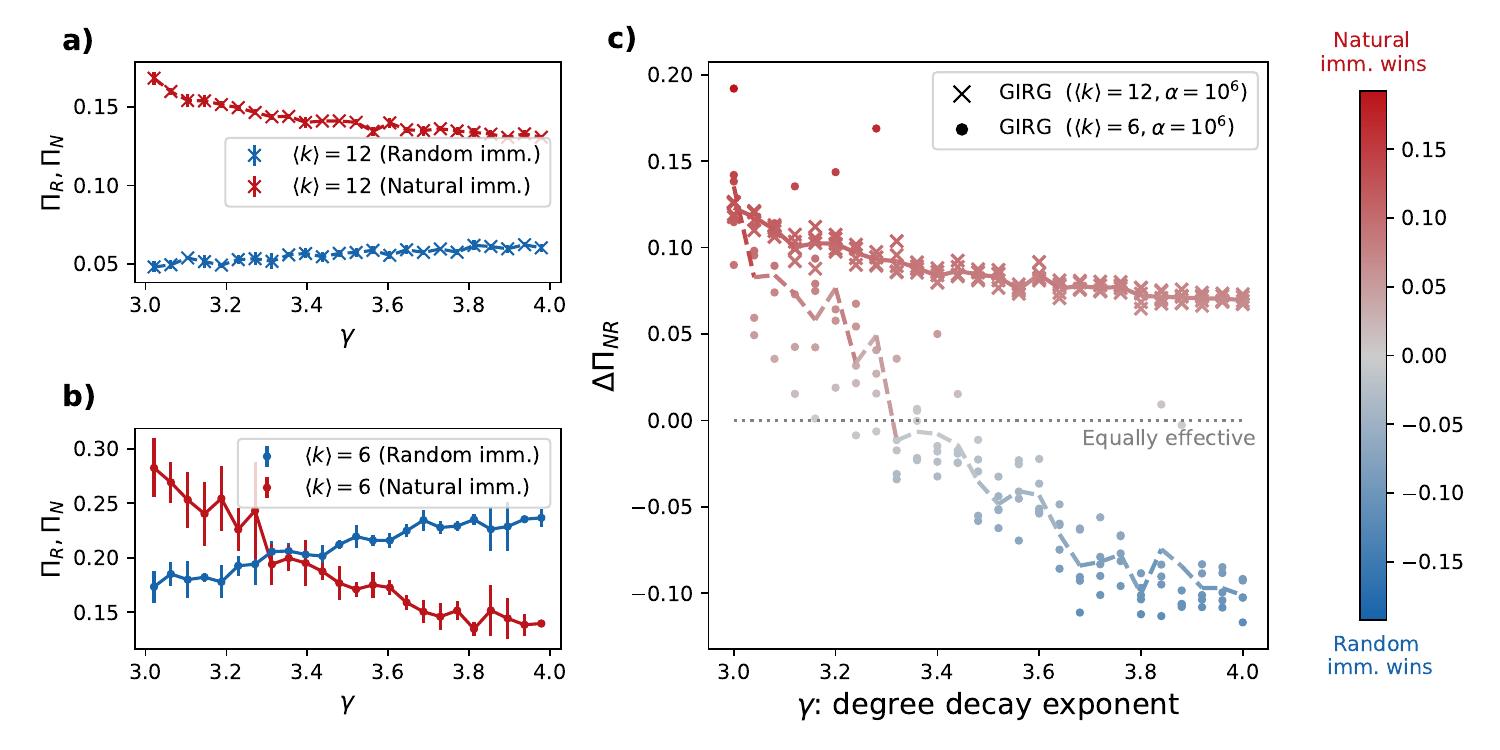}
    \caption{\textbf{Cumulative herd immunity from naturally and randomly acquired protection in strongly geometric GIRG networks.} Panels a) and b) show cumulative herd immunity from natural ($\Pi_N$, red) and random ($\Pi_R$, blue) immunization as a function of the degree-decay exponent $\gamma$ in dense ($\langle k \rangle = 12$) and sparse ($\langle k \rangle = 6$) contact networks, respectively. Networks are generated using the Geometric Inhomogeneous Random Graph (GIRG) model with $N= 10^4$ and $\alpha = 10^6$. Panel c) displays the difference between natural and random herd immunity levels (captured by $\Delta\Pi_{NR}=\Pi_N - \Pi_R$) for both the denser and sparser case, with values color-coded according to the colorbar on the right side of panel c). In panels a) and b) each point corresponds to an average over five independent network realizations. In panel c) results for individual networks are visualized as a scatter plot (points and crosses for sparse and dense networks, respectively), while color-coded lines indicate their averages. The grey dotted line denotes equality between the two strategies $\Delta\Pi_{NR}=0$.   
}
    \label{fig:effectiveness}
\end{figure}

More detailed results for geometric and non-geometric scale-free (SF) networks are shown in Fig.~\ref{fig:effectiveness} and Fig.~\ref{fig:effectiveness_2}, respectively. Both figures display the cumulative herd immunity levels induced by each immunization strategy individually as a function of the degree-decay exponent $\gamma$ (panels a,b), as well as their relative effectiveness, quantified by $\Delta\Pi_{NR}=\Pi_N-\Pi_R$ (panel c). Interestingly, as indicated by the increasing trend of $\Pi_R$ in Fig.~\ref{fig:effectiveness}a,b, random immunization becomes more effective in spatially embedded networks as $\gamma$ increases, while it decreases with $\gamma$ in the non-geometric limit (see Fig.~\ref{fig:effectiveness_2}a,b). In contrast, the performance of natural immunization is consistently weakening with increasing $\gamma$ in both settings.

Proceeding with the analysis of Fig.~\ref{fig:effectiveness}c, we can observe that for sufficiently sparse geometric SF networks and large enough $\gamma$ values, random immunization performs better than natural immunization, as indicated by negative values of $\Delta\Pi_{NR}$. This is a surprising observation, as its non-geometric counterpart consistently shows greater effectiveness for natural immunization (see Fig.~\ref{fig:effectiveness_2}c for comparison). Similar opposing behaviors in the geometric and non-geometric limits have previously been reported for networks with negative binomial degree distributions~\cite{Hiraoka2023HerdImmunity}. Furthermore, Fig.~\ref{fig:effectiveness}c also shows that although the relative performance of the two strategies depends on $\gamma$ in the sparser limit ($\left<k\right>=6$), increasing network density ultimately makes natural immunization the more effective scheme. This behavior again highlights a trend-reversal phenomenon in geometric GIRGs for $\gamma$ values slightly above $\gamma=3$.

\newpage

\section{U.S. Mobility Flow Networks}
\label{sect:safegraph}
As outlined in the main text of this article, the relative performance of natural versus random immunization in state-level mobility networks from the U.S. Mobility Flow dataset~\cite{SafeGraph2020, US_mobility_flow} is highly sensitive to the average degree induced by non-pharmaceutical interventions (see Fig.~4a in the main text). In several U.S. states -- such as California, New Jersey, and Massachusetts -- the relative effectiveness of these strategies can even flip between the pre-lockdown (6 April 2019) and lockdown periods (8 April 2020), making different schemes optimal in each phase. Additional key properties of the state-level mobility networks studied in the main text are summarized in Table~\ref{tab:state_abbrev_multicol}.

\begin{figure}[htb!]
    \centering
\includegraphics[width=1.\linewidth]{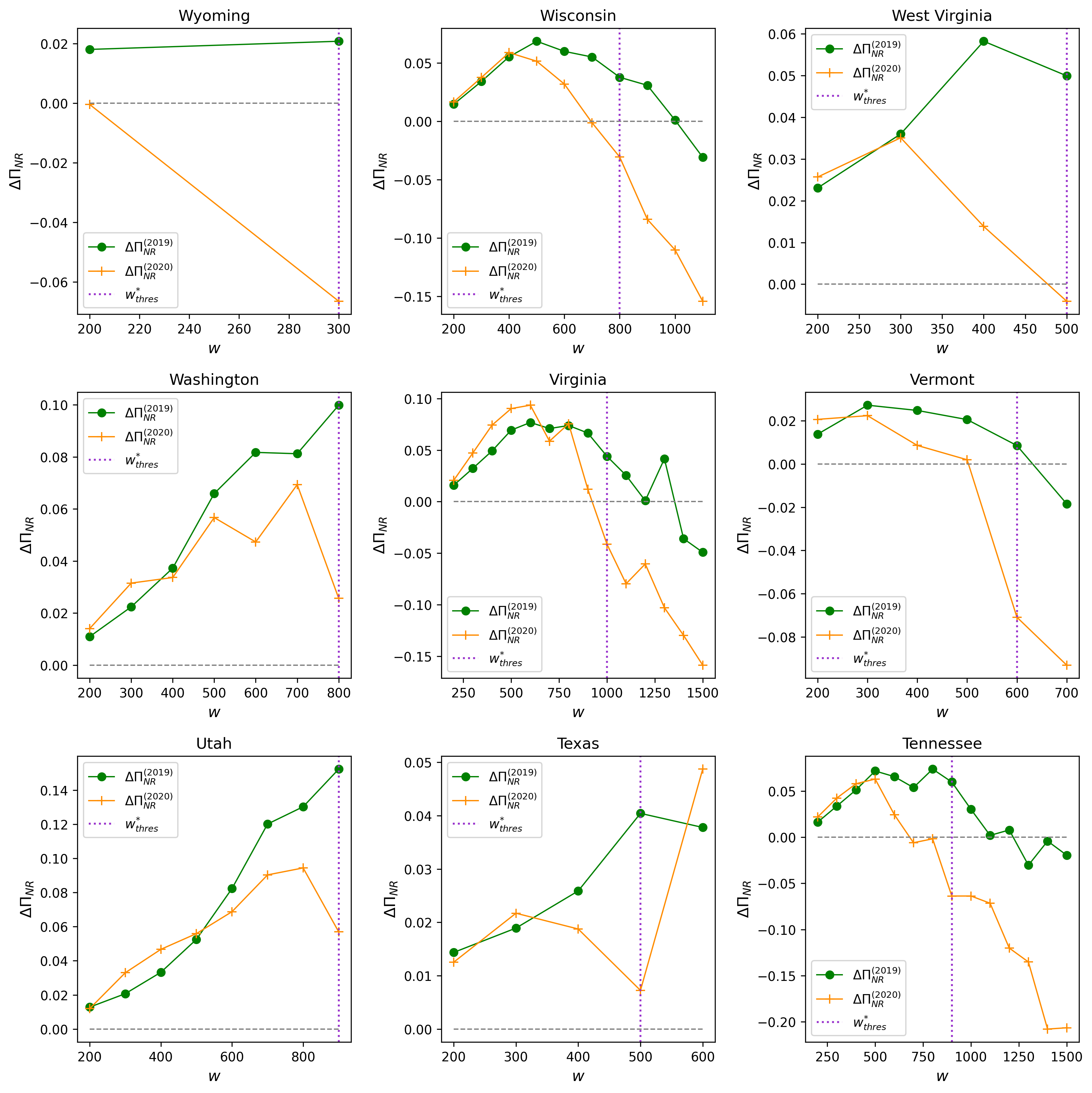}
    \caption{  \textbf{Signed difference in effectiveness between natural and random immunization strategies ($\Delta\Pi_{NR}=\Pi_N - \Pi_R$) as a function of the mobility weight threshold.} Results are shown for state-level networks---Wyoming, Wisconsin, and West Virginia (top row); Washington, Virginia, and Vermont (middle row); Utah, Texas, and Tennessee (bottom row). Orange curves represent the pre-lockdown period (6 April 2019), and green curves represent the during-lockdown period (8 April 2020). In each panel, the gray dashed line indicates equal effectiveness, while the vertical purple dotted line marks the optimal threshold value $w_{\text{thres}}$ as defined in the main text.
}
    \label{fig:state_batch1}
\end{figure}

\begin{figure}[htb!]
    \centering
\includegraphics[width=1.\linewidth]{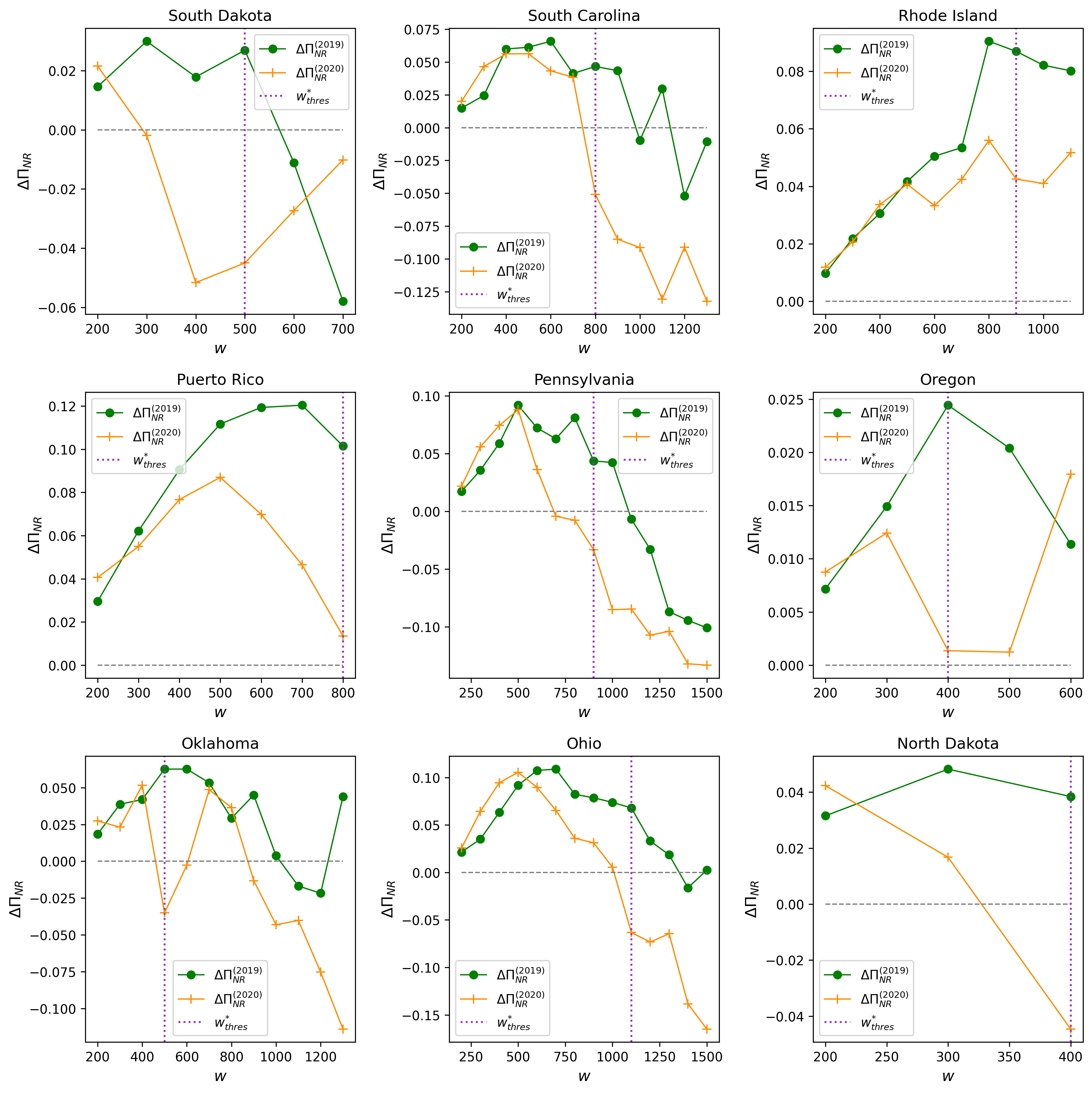}
    \caption{  \textbf{Signed difference in effectiveness between natural and random immunization strategies ($\Delta\Pi_{NR}=\Pi_N - \Pi_R$) as a function of the mobility weight threshold.} Results are shown for state-level networks---South Dakota, South Carolina, Rhode Island (top row); Puerto Rico, Pennsylvania, Oregon (middle row); Oklahoma, Ohio and North Dakota (bottom row). Orange curves represent the pre-lockdown period (6 April 2019), and green curves represent the during-lockdown period (8 April 2020). In each panel, the gray dashed line indicates equal effectiveness, while the vertical purple dotted line marks the optimal threshold value $w_{\text{thres}}$ as defined in the main text.
}
    \label{fig:state_batch2}
\end{figure}

\begin{figure}[htb!]
    \centering
\includegraphics[width=1.\linewidth]{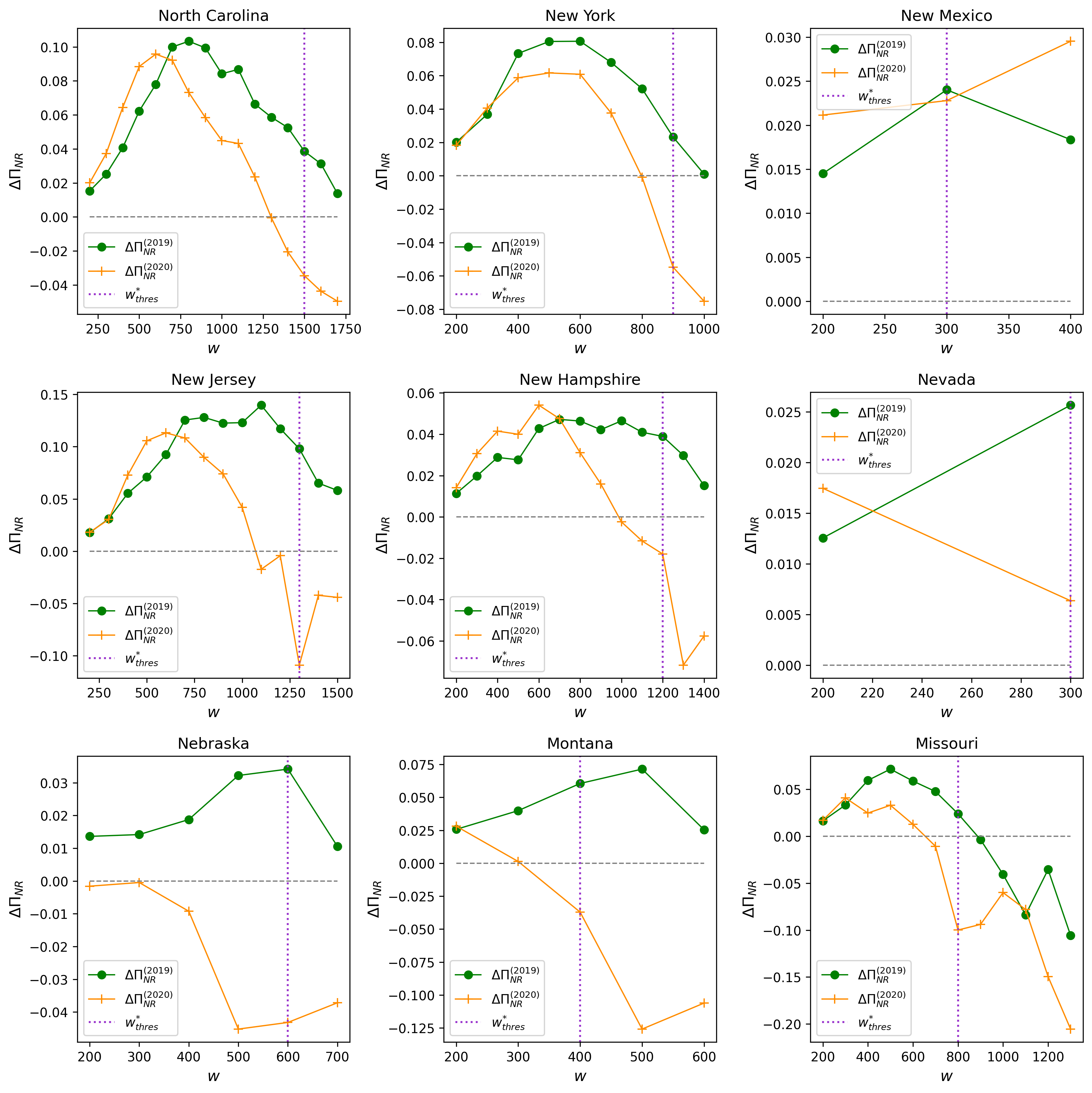}
    \caption{  \textbf{Signed difference in effectiveness between natural and random immunization strategies ($\Delta\Pi_{NR}=\Pi_N - \Pi_R$) as a function of the mobility weight threshold.} Results are shown for state-level networks---North Carolina, New York, New Mexico (top row); New Jersey, New Hampshire, Nevada (middle row); Nebraska, Montana and Missouri (bottom row). Orange curves represent the pre-lockdown period (6 April 2019), and green curves represent the during-lockdown period (8 April 2020). In each panel, the gray dashed line indicates equal effectiveness, while the vertical purple dotted line marks the optimal threshold value $w_{\text{thres}}$ as defined in the main text.
}
    \label{fig:state_batch3}
\end{figure}

\begin{figure}[htb!]
    \centering
\includegraphics[width=1.\linewidth]{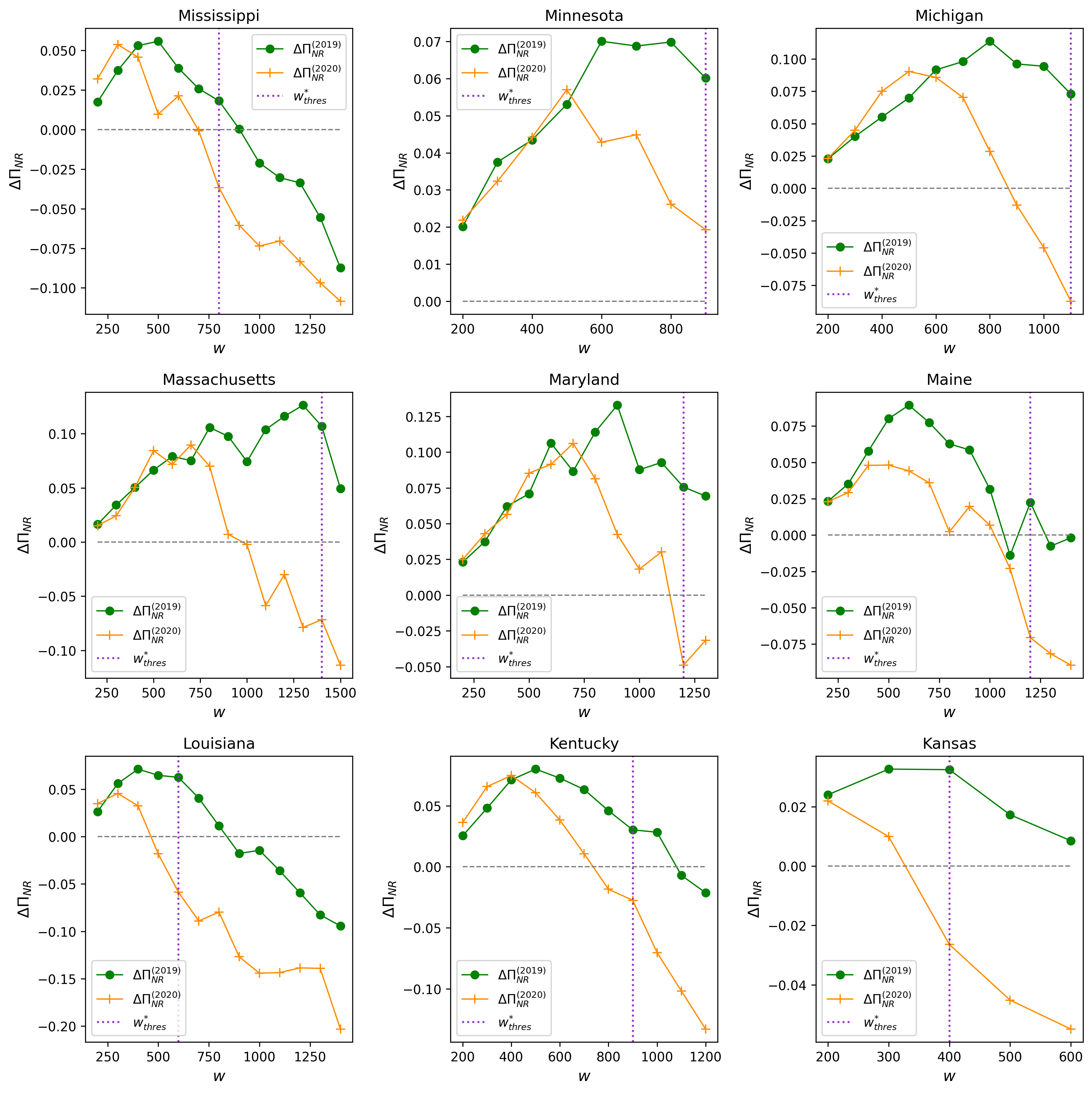}
    \caption{  \textbf{Signed difference in effectiveness between natural and random immunization strategies ($\Delta\Pi_{NR}=\Pi_N - \Pi_R$) as a function of the mobility weight threshold.} Results are shown for state-level networks---Mississippi, Minnesota, Michigan (top row); Massachusetts, Maryland, Maine (middle row); Louisiana, Kentucky and Kansas (bottom row). Orange curves represent the pre-lockdown period (6 April 2019), and green curves represent the during-lockdown period (8 April 2020). In each panel, the gray dashed line indicates equal effectiveness, while the vertical purple dotted line marks the optimal threshold value $w_{\text{thres}}$ as defined in the main text.
}
    \label{fig:state_batch4}
\end{figure}

\begin{figure}[htb!]
    \centering
\includegraphics[width=1.\linewidth]{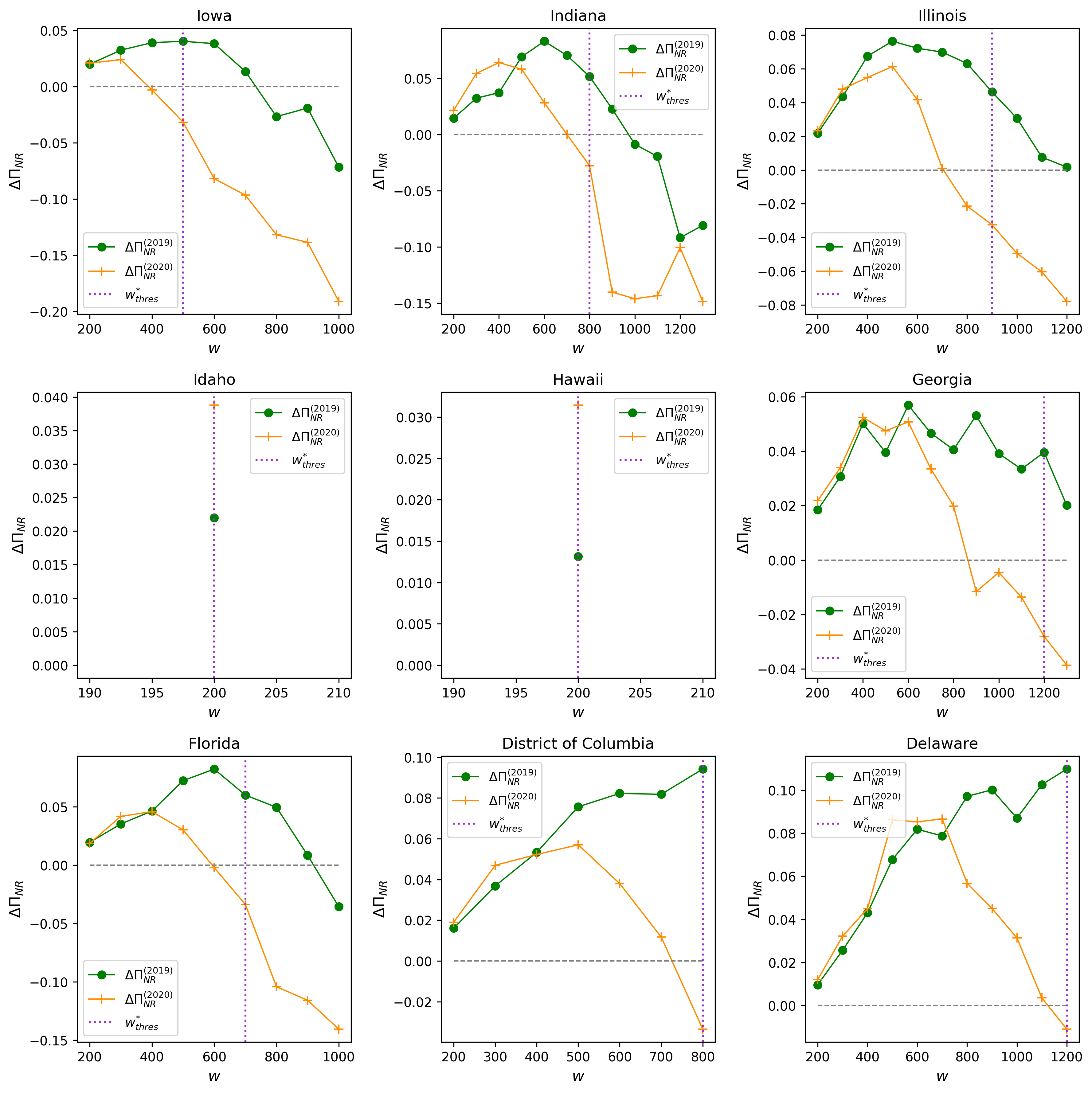}
    \caption{  \textbf{Signed difference in effectiveness between natural and random immunization strategies ($\Delta\Pi_{NR}=\Pi_N - \Pi_R$) as a function of the mobility weight threshold.} Results are shown for state-level networks---Iowa, Indiana, Illinois (top row); Idaho, Hawaii, Georgia (middle row); Florida, District of Columbia (D.C.) and Delaware (bottom row). Orange curves represent the pre-lockdown period (6 April 2019), and green curves represent the during-lockdown period (8 April 2020). In each panel, the gray dashed line indicates equal effectiveness, while the vertical purple dotted line marks the optimal threshold value $w_{\text{thres}}$ as defined in the main text.
}
    \label{fig:state_batch5}
\end{figure}

\begin{figure}[htb!]
    \centering
\includegraphics[width=1.\linewidth]{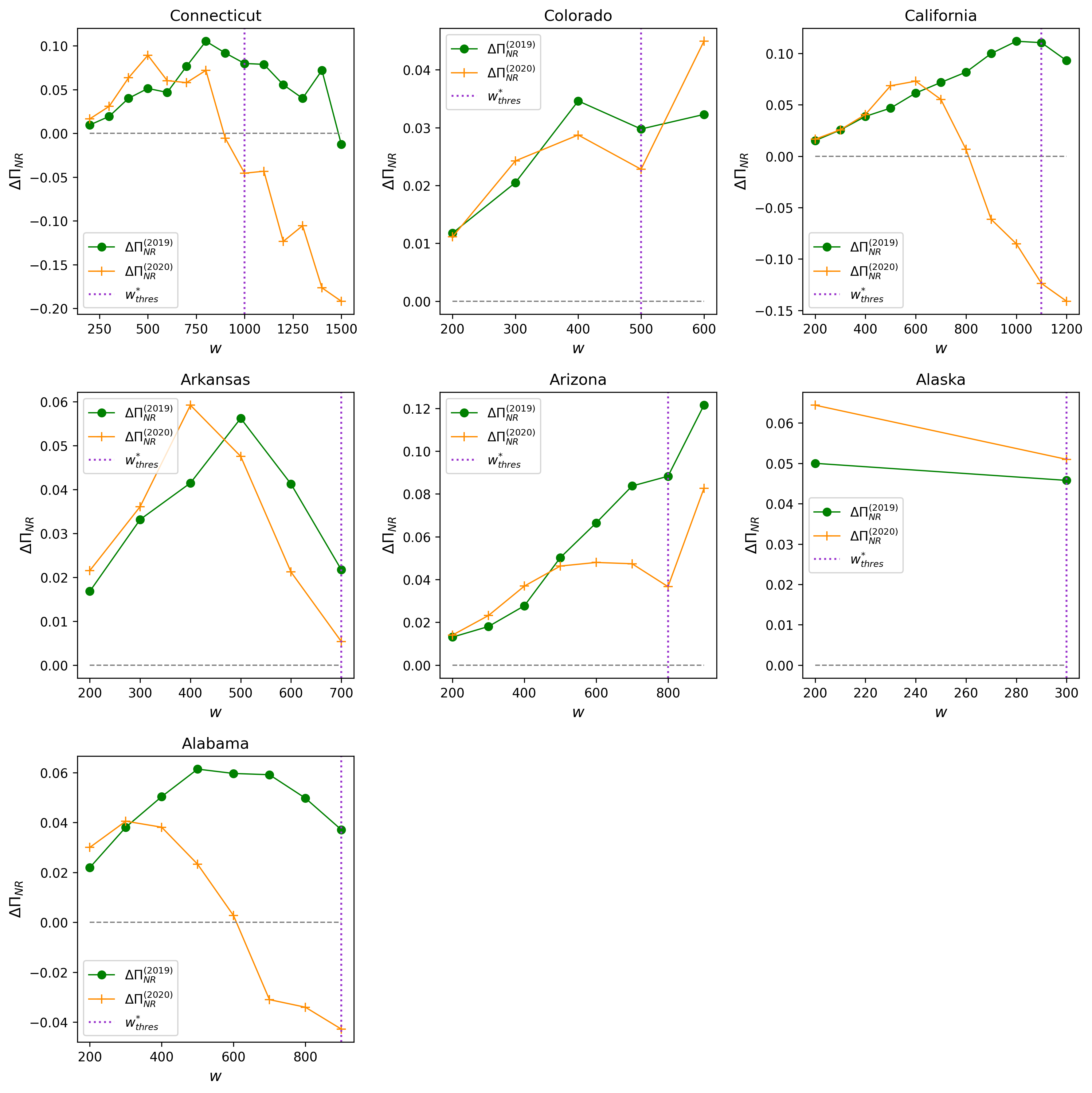}
    \caption{\textbf{Signed difference in effectiveness between natural and random immunization strategies ($\Delta\Pi_{NR}=\Pi_N - \Pi_R$) as a function of the mobility weight threshold.} Results are shown for state-level networks---Connecticut, Colorado, California (top row); Arkansas, Arizona, Alaska (middle row); Alabama (bottom row). Orange curves represent the pre-lockdown period (6 April 2019), and green curves represent the during-lockdown period (8 April 2020). In each panel, the gray dashed line indicates equal effectiveness, while the vertical purple dotted line marks the optimal threshold value $w_{\text{thres}}$ as defined in the main text.
}
    \label{fig:state_batch6}
\end{figure}

Here, we present simulation results that provide a more comprehensive assessment of the immunization strategies discussed above. Figures~\ref{fig:state_batch1}--\ref{fig:state_batch6} depict the signed difference in effectiveness between natural and random immunization as a function of the mobility weight threshold for each state-level network. The results reveal a non-trivial dependence on the choice of threshold value. Specifically, for many U.S. states, the signed difference, defined as $\Delta\Pi_{NR} = \Pi_N - \Pi_R$, displays a pronounced inverted U-shaped trend in both the pre-lockdown and lockdown periods, indicating the existence of a threshold at which the performance gap between the two strategies reaches its maximum. Particularly notable examples are Wisconsin and Virginia in Fig.~\ref{fig:state_batch1}, Ohio in Fig.~\ref{fig:state_batch2}, North Carolina and New York in Fig.~\ref{fig:state_batch3}, Louisiana and Kentucky in Fig.~\ref{fig:state_batch4}, Illinois and Florida in Fig.~\ref{fig:state_batch5} or Alabama in Fig.~\ref{fig:state_batch6}. In the subsequent sections, we turn to examine the degree specificity of the strategy-reversal phenomenon, as well as its dependence on the applied mobility weight threshold.  

\subsection{Degree specificity in the strategy reversal phenomenon}

\begin{figure}[htb!]
    \centering
\includegraphics[width=.6\linewidth]{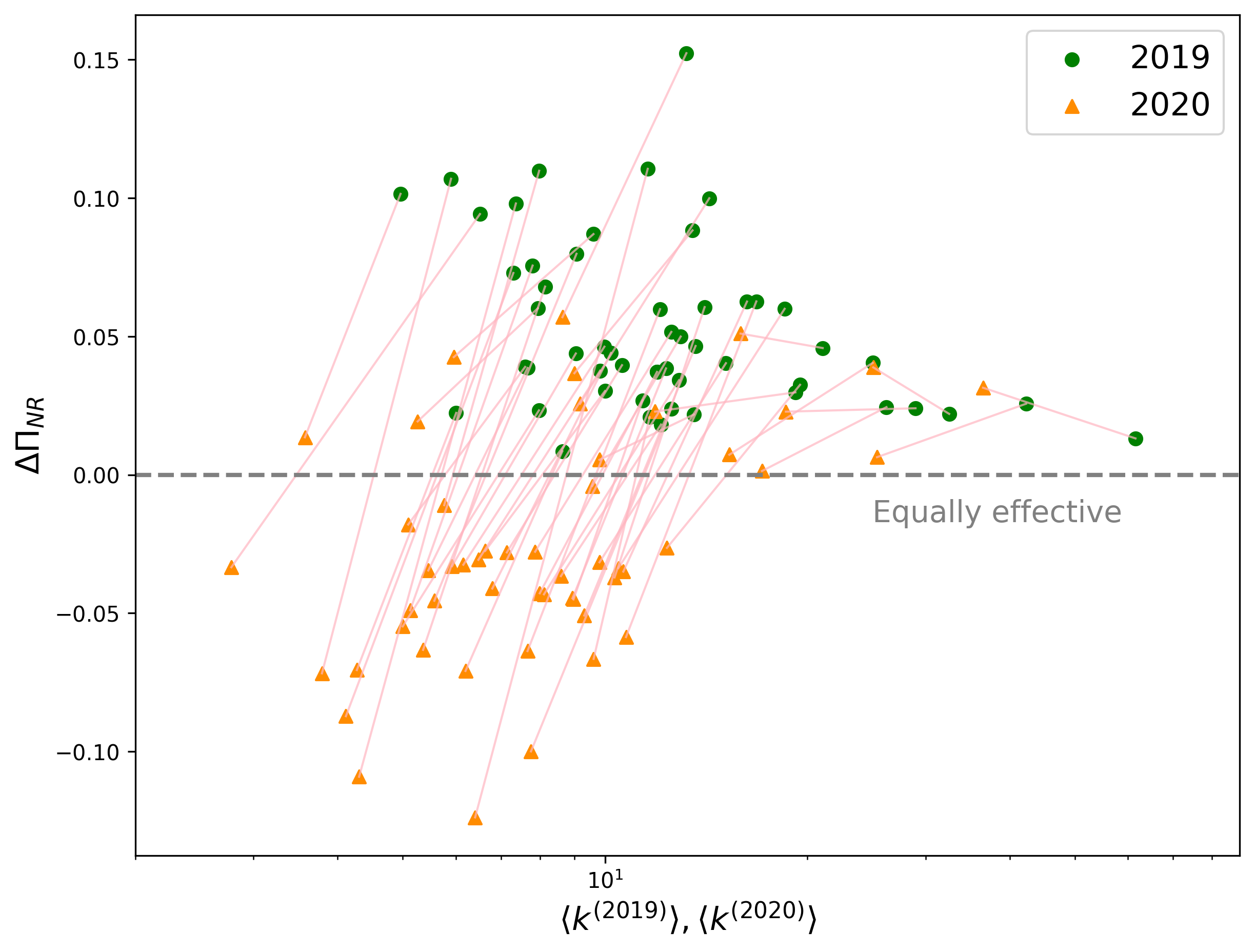}
    \caption{\textbf{Signed differences in the effectiveness of natural versus random immunization strategies before and during lockdown, shown as a function of the corresponding average degree pairs.} Green circles indicate the signed differences $\Delta\Pi_{NR}$ measured in the pre-lockdown period, while orange triangles represent the corresponding values during lockdown. Pink solid lines connect the two phases of the same U.S. state; their horizontal positions reflect the state’s average degree before and during the lockdown period, denoted by $\langle k^{(2019)}\rangle$ and $\langle k^{(2020)}\rangle$, respectively. The dashed grey line marks the threshold of equal effectiveness, where natural and random immunization perform identically under the chosen metric.}
\label{fig:scatter_k_pi_wopt}
\end{figure}

In our analysis, the mobility-weight threshold $w_{\text{thres}}$ functions as a control parameter of the network. Adjusting this threshold directly tunes the resulting average degree, which in turn determines the size of the largest connected component and, consequently, the network’s criticality. As explained in the Methods section, the upper limit for the threshold value $w^{\text{max}}_{\text{thres}}$ is selected separately for each state in such a way that the corresponding networks remain highly connected during the lockdown phase (April 6th, 2020). For a more detailed description, see the Methods section. As a result of the above, $w^{\text{max}}_{\text{thres}}$ approximately marks the onset of criticality in the during-lockdown period, whereas the lower threshold limit, which is always set to $w^{\text{min}}_{\text{thres}}=200$ produces substantially denser networks that are deeply within the supercritical regime. The optimal threshold $ w^{*}_{\text{thres}}$, as defined in the Methods section and reported in Table~\ref{tab:state_abbrev_multicol}, lies between these two extremes. The simulation results corresponding to the optimal choice $w^{*}_{\text{thres}}$ are shown in Fig.~4a-c of the main text. Complementing this analysis, Fig.~\ref{fig:scatter_k_pi_wopt} illustrates, for all U.S. states, how the strategy-reversal phenomenon depends on the relationship between the average network degrees before and during the lockdown period. The figure shows the relative effectiveness of the two strategies, quantified by $\Delta\Pi_{NR}$ in both periods, and uses a dumbbell plot to compare the average degrees before and during lockdown. Green circles denote the pre-lockdown period, while orange triangles represent the lockdown period. Pink solid lines connect the two observations for each state; if a line crosses the line of equal effectiveness (indicated by the dashed gray line), the corresponding state exhibits a strategy reversal.

Although the average degrees vary across states, in agreement with Fig.~4a-c we can observe that the majority of states display the strategy-reversal phenomenon, though not all do. The largely consistent upright directionality of the pink dumbbells in Fig.~\ref{fig:scatter_k_pi_wopt} suggests that mobility restrictions acted primarily in two ways: they reduced the average degree of the mobility networks while simultaneously increasing the relative effectiveness of random immunization. For most states, the reduction in average degree during the lockdown period was sufficient to induce the observed strategy reversal; however, in states where the average degree remained comparatively high, no reversal occurred.

Nevertheless, although the reversal phenomena clearly influenced by a density drop in the network, it cannot be directly associated with a specific degree regime; dumbbells in Fig.~\ref{fig:scatter_k_pi_wopt} crossing the equality line (grey dashed line) are distributed across a wide span of average degrees. Conversely, states that do not exhibit a reversal are also spread across the degree spectrum, occurring both at relatively high average degrees (above $10$) and at lower values (around $3$–$5$). This indicates that a variety of factors can determine whether the phenomenon emerges in real-world networks: some networks behave more similarly to configuration model networks and do not exhibit the phenomenon regardless of their average degree, while others, e.g. those showing strong signs of geometric correlations, do display strategy reversal. Crucially, this indicates that the strategy-reversal phenomenon does not arise only because some states transitioned from denser mobility networks in 2019 to sparser ones in 2020 compared to states that do not show a reversal. Rather, it showcases how subtle, non-trivial intervention-induced changes in network structure --simultaneously affecting degree, spatial embeddedness, and other inherent structural properties -- shape the effectiveness of immunization strategies.

\subsection{Dependence of the strategy reversal phenomenon on the mobility weight threshold}

\begin{figure}[htb!]
    \centering
\includegraphics[width=.8\linewidth]{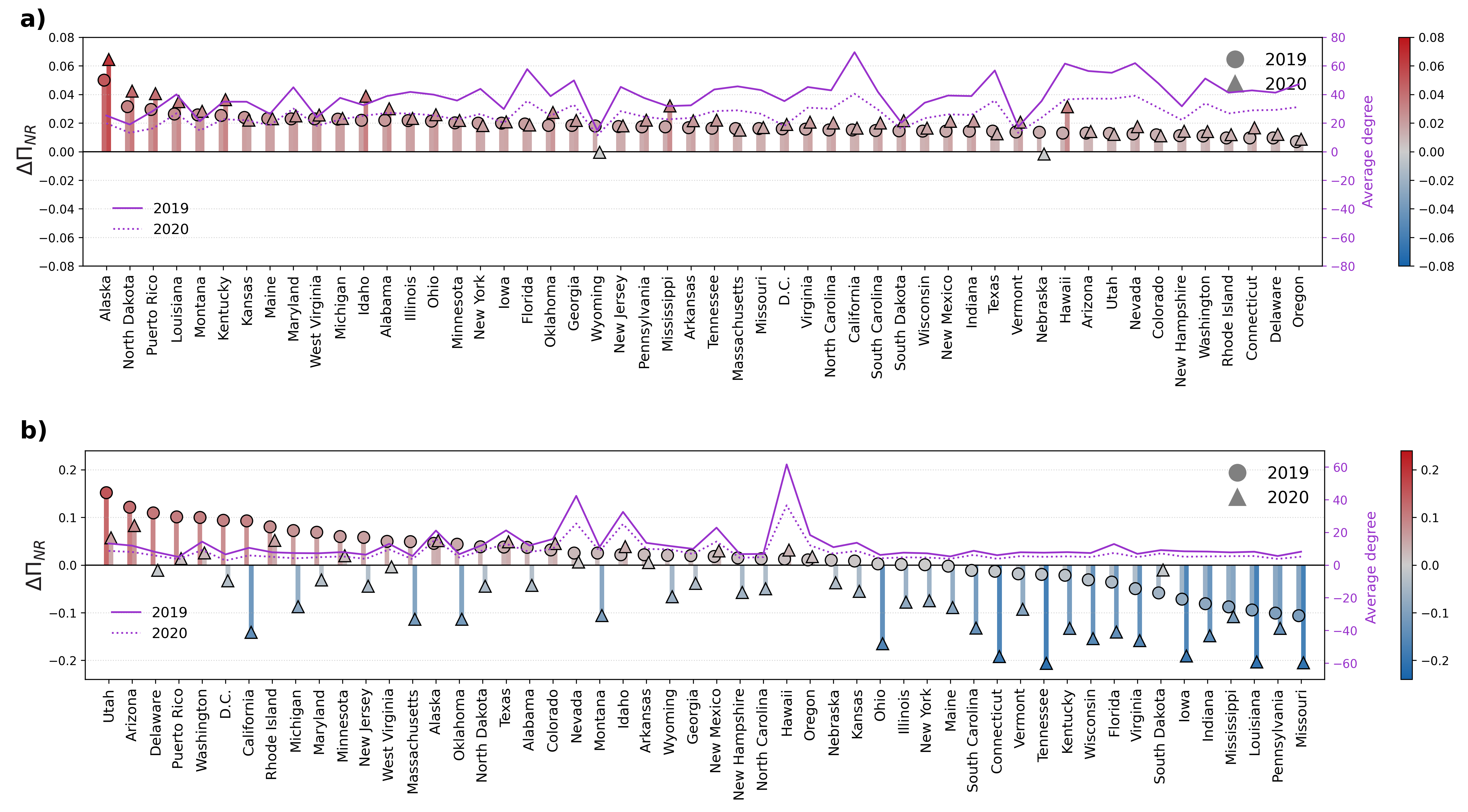}
    \caption{\textbf{Relative effectiveness of natural versus random immunization on U.S. state-level mobility networks before and during lockdown periods.} Panel (a) shows a bar chart of the differences in effectiveness between natural and random immunity for each state, with circles representing the pre-lockdown period and triangles representing the during-lockdown period. Solid and dotted purple lines indicate the average degrees of the 2019 and 2020 state-level networks, respectively. A minimum weight threshold of $w^{\text{min}}_{\text{thres}}=200$ was applied in panel (a). Panel (b) presents a similar bar chart, but using the maximum weight threshold $w^{\text{max}}_{\text{thres}}$ for each state-level network.}
    \label{fig:map_var}
\end{figure}

Figure~\ref{fig:map_var} shows how the strategy reversal phenomenon changes as the mobility-weight threshold and, consequently, the degree of criticality varies across the state-level networks. Panel~\ref{fig:map_var}a summarizes the results for strongly supercritical (dense) networks, where a uniform fixed mobility threshold of $w^{\text{min}}_{\text{thres}} = 200$ is applied to all states. In this regime, no trend reversal is observed; natural (disease-induced) immunity is consistently more effective than random immunization.

In contrast, panel~\ref{fig:map_var}b presents results in which thresholds are again state-specific but chosen to push each network as close to criticality as possible; that is, $w^{\text{max}}_{\text{thres}}$ is applied as the highest threshold that still preserves connectivity in each state. Such threshold values are summarized in Table~\ref{tab:state_abbrev_multicol}. Under these conditions, the trend-reversal phenomenon re-emerges, though with new details. Several states, such as California and Michigan, exhibit a clear strategy reversal near criticality; others, including Utah and Arizona, show stronger performance for natural immunization irrespective of interventions; and yet others, such as Missouri, Pennsylvania, and Ohio, along with several additional states in the rightmost region of Fig.~4b consistently display greater effectiveness for uniform vaccination in both periods.

\scriptsize
\begin{longtable}{l  l  c c c c c c c c c}
\hline
State & Abbrev. & $w^{\text{min}}_{\text{thres}}$ & $w^{\text{max}}_{\text{thres}}$ & $w^{*}_{\text{thres}}$ 
& $N^{(2019)}$ & $N^{(2020)}$ & $E^{(2019)}$ & $E^{(2020)}$ 
& $\langle k^{(2019)}\rangle$ & $\langle k^{(2020)}\rangle$ \\
\hline
\endfirsthead
\hline
State & Abbr. & $w^{\text{min}}_{\text{thres}}$ & $w^{\text{max}}_{\text{thres}}$ & $w^{*}_{\text{thres}}$ 
& $N^{(2019)}$ & $N^{(2020)}$ & $E^{(2019)}$ & $E^{(2020)}$ 
& $\langle k^{(2019)}\rangle$ & $\langle k^{(2020)}\rangle$ \\
\hline
\hline
\endhead
\endfoot
Alabama & AL & 200 & 900 & 900 & 1151 & 1130 & 6877 & 4518 & 11.95 & 8.0 \\
Alaska & AK & 200 & 300 & 300 & 152 & 136 & 1602 & 1082 & 21.08 & 15.91 \\
Arizona & AZ & 200 & 900 & 800 & 1487 & 1476 & 10035 & 6645 & 13.5 & 9.0 \\
Arkansas & AR & 200 & 700 & 700 & 681 & 680 & 4615 & 3336 & 13.55 & 9.81 \\
California & CA & 200 & 1200 & 1100 & 7711 & 7293 & 44631 & 23370 & 11.58 & 6.41 \\
Colorado & CO & 200 & 600 & 500 & 1233 & 1227 & 11836 & 7284 & 19.2 & 11.87 \\
Connecticut & CT & 200 & 1500 & 1000 & 806 & 772 & 3655 & 2151 & 9.07 & 5.57 \\
Delaware & DE & 200 & 1200 & 1200 & 210 & 196 & 838 & 565 & 7.98 & 5.77 \\
District of Columbia & DC & 200 & 800 & 800 & 167 & 141 & 544 & 196 & 6.51 & 2.78 \\
Florida & FL & 200 & 1000 & 700 & 4113 & 4077 & 38054 & 21338 & 18.5 & 10.47 \\ 
Georgia & GA & 200 & 1300 & 1200 & 1879 & 1837 & 9958 & 6562 & 10.6 & 7.14 \\
Hawaii & HI & 200 & 200 & 200 & 314 & 310 & 9664 & 5660 & 61.55 & 36.52 \\
Idaho & ID & 200 & 200 & 200 & 297 & 297 & 4828 & 3723 & 32.51 & 25.07 \\
Illinois & IL & 200 & 1200 & 900 & 2860 & 2769 & 14271 & 8522 & 9.98 & 6.16 \\
Indiana & IN & 200 & 1300 & 800 & 1475 & 1449 & 9256 & 5703 & 12.55 & 7.87 \\
Iowa & IA & 200 & 1000 & 500 & 824 & 819 & 6234 & 4019 & 15.13 & 9.81 \\
Kansas & KS & 200 & 600 & 400 & 759 & 753 & 7403 & 4655 & 19.51 & 12.36 \\
Kentucky & KY & 200 & 1200 & 900 & 1096 & 1070 & 5479 & 3550 & 10.0 & 6.64 \\
Louisiana & LA & 200 & 1400 & 600 & 1103 & 1064 & 9262 & 5717 & 16.79 & 10.75 \\
Maine & ME & 200 & 1400 & 1200 & 338 & 329 & 1015 & 703 & 6.01 & 4.27 \\
Maryland & MD & 200 & 1300 & 1200 & 1270 & 1217 & 4955 & 3121 & 7.8 & 5.13 \\
Massachusetts & MA & 200 & 1500 & 1400 & 1350 & 1240 & 3983 & 2351 & 5.9 & 3.79 \\
Michigan & MI & 200 & 1100 & 1100 & 2397 & 2252 & 8755 & 4635 & 7.3 & 4.12 \\
Minnesota & MN & 200 & 900 & 900 & 1286 & 1235 & 5108 & 3250 & 7.94 & 5.26 \\
Mississippi & MS & 200 & 1400 & 800 & 646 & 642 & 3913 & 2760 & 12.11 & 8.6 \\
Missouri & MO & 200 & 1300 & 800 & 1324 & 1322 & 8305 & 5129 & 12.55 & 7.76 \\
Montana & MT & 200 & 600 & 400 & 267 & 265 & 1878 & 1369 & 14.07 & 10.33 \\
Nebraska & NE & 200 & 700 & 600 & 516 & 514 & 3324 & 2085 & 12.88 & 8.11 \\
Nevada & NV & 200 & 300 & 300 & 682 & 680 & 14430 & 8635 & 42.32 & 25.4 \\
New Hampshire & NH & 200 & 1400 & 1200 & 289 & 283 & 1099 & 721 & 7.61 & 5.1 \\
New Jersey & NJ & 200 & 1500 & 1300 & 1863 & 1714 & 6864 & 3691 & 7.37 & 4.31 \\
New Mexico & NM & 200 & 400 & 300 & 497 & 498 & 7199 & 4627 & 28.97 & 18.58 \\
New York & NY & 200 & 1000 & 900 & 4513 & 4215 & 17993 & 10539 & 7.97 & 5.0 \\
North Carolina & NC & 200 & 1700 & 1500 & 2084 & 2032 & 7992 & 5546 & 7.67 & 5.46 \\
North Dakota & ND & 200 & 400 & 400 & 204 & 201 & 1257 & 899 & 12.32 & 8.95 \\
Ohio & OH & 200 & 1500 & 1100 & 2582 & 2497 & 10513 & 6695 & 8.14 & 5.36 \\
Oklahoma & OK & 200 & 1300 & 500 & 1033 & 1021 & 8401 & 5436 & 16.27 & 10.65 \\
Oregon & OR & 200 & 600 & 400 & 825 & 823 & 10819 & 7046 & 26.23 & 17.12 \\
Pennsylvania & PA & 200 & 1500 & 900 & 3112 & 2991 & 14079 & 8855 & 9.05 & 5.92 \\
Puerto Rico & PR & 200 & 800 & 800 & 762 & 742 & 1891 & 1327 & 4.96 & 3.58 \\
Rhode Island & RI & 200 & 1100 & 900 & 237 & 232 & 1139 & 691 & 9.61 & 5.96 \\
South Carolina & SC & 200 & 1300 & 800 & 1076 & 1062 & 7337 & 4946 & 13.64 & 9.31 \\
South Dakota & SD & 200 & 700 & 500 & 219 & 219 & 1246 & 983 & 11.38 & 8.98 \\
Tennessee & TN & 200 & 1500 & 900 & 1450 & 1438 & 8753 & 5514 & 12.07 & 7.67 \\
Texas & TX & 200 & 600 & 500 & 5208 & 5199 & 65165 & 39802 & 25.02 & 15.31 \\
Utah & UT & 200 & 900 & 900 & 573 & 568 & 3786 & 2455 & 13.21 & 8.64 \\
Vermont & VT & 200 & 700 & 600 & 181 & 179 & 782 & 555 & 8.64 & 6.2 \\
Virginia & VA & 200 & 1500 & 1000 & 1815 & 1787 & 9261 & 6073 & 10.2 & 6.8 \\
Washington & WA & 200 & 800 & 800 & 1430 & 1423 & 10220 & 6534 & 14.29 & 9.18 \\
West Virginia & WV & 200 & 500 & 500 & 490 & 485 & 3175 & 2321 & 12.96 & 9.57 \\
Wisconsin & WI & 200 & 1100 & 800 & 1310 & 1292 & 6440 & 4187 & 9.83 & 6.48 \\
Wyoming & WY & 200 & 300 & 300 & 132 & 131 & 769 & 630 & 11.65 & 9.62 \\
\hline
\caption{
\textbf{U.S. states and territories with their abbreviations, as used in this article, together with basic properties of the corresponding state-level networks.}
The quantities $w^{\min}_{\text{thres}}, w^{\max}_{\text{thres}}, w^{*}_{\text{thres}}$ denote the minimum, maximum, and optimally selected mobility thresholds.
$N^{(2019)}$ and $N^{(2020)}$ represent the numbers of nodes in the 2019 and 2020 state-level networks, respectively, while
$E^{(2019)}$ and $E^{(2020)}$ represent the corresponding numbers of edges.
}
\label{tab:state_abbrev_multicol}
\end{longtable}
\normalsize

\newpage

\newpage

\newpage

\newpage

\newpage

\newpage

\newpage

\end{document}